\documentclass[manuscript]{aastex}

\usepackage{emulateapj5}
\usepackage{epsf}
\usepackage{graphics}

\newenvironment{inlinefigure}{
\def\@captype{figure}
\noindent\begin{minipage}{0.999\linewidth}\begin{center}}
{\end{center}\end{minipage}\smallskip}

\shorttitle{Rapid and Early Galaxy Merging}
\shortauthors{C.J. Conselice}

\def\deg{$^{\circ}\,$}
\def\solm{M$_{\odot}\,$}

\begin{document}

\title{Early and Rapid Merging as a Formation Mechanism of Massive Galaxies:
Empirical Constraints}

\author{Christopher J. Conselice$^{1,2}$}

\altaffiltext{1}{California Institute of Technology, Mail Code 105-24, Pasadena
CA 91125}
\altaffiltext{2}{National Science Foundation Astronomy \& Astrophysics Fellow}

\begin{abstract}
 
We present the results of a series of empirical computations regarding
the role of major mergers in forming
the stellar masses of modern galaxies.   We
base these results on measurements of galaxy merger and star formation 
histories from $z \sim 0.5-3$.  
We re-construct the merger history of normal field
galaxies from $z \sim 3$ to $z \sim 0$ as a function of initial
stellar mass using published pair fractions and merger fractions from
structural analyses.  We calibrate the observed
merger time-scale and mass ratios for galaxy mergers
using self-consistent N-body models of mergers, composed of dark matter and 
stars, with mass ratios from 1:1 to 1:5 with various orbital
properties and viewing angles.   We use these
simulations to determine the time-scales and mass ratios that
produce structures that would be identified as major mergers.
Based on these calculations we argue that a typical massive
galaxy at $z \sim 3$ with M$_{*} > 10^{10}$ \solm 
undergoes 4.4$^{+1.6}_{-0.9}$ major mergers at $z > 1$, with no further
mergers at $z < 1$..
We find that by $z \sim 1.5$ the stellar mass of an average massive 
galaxy is relatively established, a scenario qualitatively favored in a 
$\Lambda$ dominated universe, and independently suggested through 
stellar population analyses.   Through empirical measurements of
the star formation induced in these mergers, and using
assumptions concerning the star formation history, we
argue that the final stellar masses of these systems increases by
as much as a factor of $\sim 100$ allowing Lyman-break
galaxies, which tend to have low stellar masses,
to become the most massive  galaxies in today's
universe with M $>$ M$^{*}$.  Induced star formation 
accounts for 10-30\% of the stellar mass formed in these
galaxies at $z < 3$.   A comparison to semi-analytic
models of galaxy formation shows that Cold Dark Matter (CDM) models 
consistently under-predict the merger fraction, and rate of merging, of massive
galaxies at high redshift.  This suggests that massive galaxy
formation occurs through more merging than predicted in CDM models, 
rather than a rapid early collapse.  Finally, we argue that rapid and early 
merging as the formation mechanisms of
massive galaxies naturally explains many previously paradoxical properties
of elliptical/massive galaxies.

\end{abstract}

\section{Introduction}

The prime motivation for studying galaxies at high redshift is 
to learn the history and physics responsible for their formation.  
The first step in this process is to identify
galaxy populations in the distant universe. This has been accomplished
to a degree with the discovery and systematic study of sub-mm 
sources (Blain et al. 2002), Lyman-$\alpha$ emitters (Kodaira et al.
2003), extremely red
objects (Moustakas et al. 2004) and  Lyman-Break galaxies 
(LBGs; Giavalisco 2002).  The latter category are UV bright 
star forming galaxies, typically at $z > 2.5$, which are thus far the best 
studied high redshift galaxy type.    How these different galaxy
populations are related to each other, and to galaxies in the modern universe
is a major area of research, and is still largely an open question.   Another
related issue is the history of the physical processes
that drive galaxy formation, such as mergers, gas cooling or some
combination of these (Conselice et al. 2001a). One fundamental
problem is whether or not we have identified the progenitors of the most
massive galaxies in the local universe at these early times.  
There are several indications that at least some 
massive galaxy progenitors are high redshift LBGs. This includes
clustering statistics showing that LBGs inhabit massive dark
matter halos (e.g., Giavalisco et al. 1998; Adelberger et al. 1998; Adelberger
et al. 2005).  Other,
possibly more massive and evolved, galaxies exist at these redshifts,
identified through infrared and sub-mm surveys, which may or may not
be a subset of the LBG population 
(e.g., Franx et al. 2003; Chapman et al. 2003; Daddi et al. 2004).  However, 
simple calculations show that the most massive galaxies today should be 
forming stars at $z \sim 3$, and it seems likely that a sub-set of the LBG
population are in some form the
progenitors of a sub-set of modern massive galaxies.

There is however a problem with interpreting LBGs 
as the high redshift progenitors of modern massive galaxies. 
LBGs are nearly as luminous as the most luminous galaxies in the nearby
universe, yet they have stellar populations with low mass to light
ratios (e.g., Papovich et al. 2001; Shapely et al. 2001), resulting in 
relatively low computed stellar masses.   In fact, very few LBGs have stellar 
masses larger than the local value of M$^{*}$, the characteristic galaxy mass 
(Papovich et al. 2001; Shapley et al. 2001).
These low mass to light ratios imply that there is considerable
star formation in LBGs, thus it is possible that extended star formation will
produce galaxies with a large stellar mass by $z \sim 0$. However, when the
observed ongoing
star formation of LBGs is evolved to the present time, using reasonable
assumptions, the resulting stellar masses are generally not large enough to be 
representative of massive galaxies in the nearby
universe (Papovich et al. 2001).    For LBGs to produce a large
enough stellar mass by the time they evolve to $z \sim 0$, they 
must have multiple episodes of star formation, or
somehow prolong or rejuvenate starbursts.
Until recently, there has been no obvious way to determine if or when 
Lyman-break galaxies will undergo future bursts.

One possible way for LBGs to
increase their stellar mass and evolve into modern
ellipticals, or spiral bulges, is through the merger process (e.g., Conselice
2003a; Nagamine et al. 2005). Major 
mergers\footnote{A major merger is defined as a merger between two galaxies 
with a mass ratio 1:3 or lower. A minor merger is between systems with
a mass ratio higher than 1:3.}
are efficient at triggering star formation as well as building up the stellar
mass of a galaxy through the addition of mass from the ``accreted''
galaxy.   The idea that LBGs are involved in major mergers has been suggested
through theoretical and empirical arguments (e.g, Somerville et al. 2001;
Conselice et al. 2003a; Weatherley \& Warren 2003), yet no detailed 
calculations have been performed to determine the number of mergers these
systems go through, or how much mass is potentially 
added to LBGs due to this process.  

We examine the evidence for major mergers at high redshift
in this paper and calculate,
based on empirical data, how many mergers a typical $z \sim 3$ galaxy 
undergoes as a function of redshift, finding that a typical
massive galaxy undergoes on average 4.4$^{+1.6}_{-0.9}$ major galaxy
mergers. This is an extension of our earlier work where the merger fraction
is derived utilizing deep HST imaging of the Hubble Deep Field (Conselice
et al. 2003). This new paper presents a reanalysis of these merger
fractions and includes N-body models to calibrate time-scales in which
galaxies would be seen as mergers through their structures.  This allows
us to produce the first galaxy merger rates and merger histories at $z > 1$.
We furthermore use the observed stellar masses
of galaxies undergoing major mergers to determine the average stellar mass 
accumulation due to mergers, as a function of mass, and
as a function of time.  We also derive the amount of star
formation likely induced by these mergers.   We conclude that
due to the merger process LBGs can become the most massive galaxies in
today's universe through the major merger process, and most of this formation
is complete by $z \sim 1.5$.  We compare our empirical predictions with 
observations of how the stellar masses of galaxies grow with time, and with
specific predictions from Cold Dark Matter based models.

Throughout this paper we use the following cosmology: H$_{0}$ = 70 km s$^{-1}$
Mpc$^{-1}$, 
$\Omega_{\Lambda}$ = 0.7 and $\Omega_{\rm m}$ = 0.3. 
\S 2 summarizes the various data used in this paper to perform the 
calculations, \S 3 is an analysis of merger fractions and rates, and their
evolution, including a new fitting formalism, \S 4 contains a
calculation of how many mergers a high redshift galaxy will undergo, and how
much stellar mass is added during this process, while \S 5 is a discussion
of our results, and \S 6 is a summary.

\section{Data and Methods}

To constrain the role of mergers in forming galaxies requires 
understanding the merger history through pair and structural methods, and the 
measured stellar masses of $z \sim 3$ galaxies.
These data come from many sources, as we require
constraints on the merger and mass assembly history of galaxies out to 
$z \sim 3$.   Merger fractions and histories derived using
pairs of galaxies (\S 3.1) come from 
Patton et al. (2000), Le F\'{e}vre et al. (2000), Patton et al. (2002)
and Lin et al. (2005) at $z < 1$.   We use
previously published and cataloged data from Conselice et al. (2003a) and
Conselice, Blackburne \& Papovich (2005) for galaxy mergers  
at $z > 0.5$ that utilize galaxy structural information
from the Hubble Deep Field North and South.  We also
use the stellar masses calculated and tabulated for high redshift ($z > 2$)
galaxies to constrain the initial masses for these
mass growth calculations (e.g., Papovich et al. 2001).

We present merger fractions as a function of redshift for galaxies
at $z > 1$ based on galaxies seen in the
 Hubble Deep Field North (HDF-N) and South (HDF-S).
We identify mergers in Conselice et al. (2003a)
and in this paper through the CAS (Concentration, Asymmetry, Clumpiness) 
system (Conselice 2003b; Conselice, Gallagher \& Wyse 2002).   
The basic idea behind CAS morphologies 
is that galaxies undergoing major mergers can be identified through their large
asymmetries measured in the rest-frame optical (cf. Windhorst et al. 2002;
Papovich et al. 2003 for the UV). The merger
fractions for galaxies in the HDF-N up to $z \sim 3$ are discussed in
detail in Conselice et al. (2003a).  We present a new analysis
here using data from both the HDF-N and HDF-S, particularly through deep
NICMOS imaging of the HDF-N for high$-z$ systems in \S 3.1.4 
(also see Dickinson et al. 2000).   

\section{Analysis}

\subsection{Merger Fraction Evolution}

\subsubsection{Definitions}

We define the galaxy merger fraction $f_{\rm gm}$ as the number of galaxies 
undergoing a merger ($N_{\rm gm}$) 
divided by the total number of galaxies  ($N_{\rm T}$) 
within a given redshift ($z$), stellar mass (M$_{*}$) or ($\vee$)
luminosity (M$_{\rm B}$) range,

\begin{equation}
f_{\rm gm} (z, {\rm M_{B}} \vee {\rm M_{\star}})= \frac{N_{\rm gm}}{N_{\rm T}}.
\end{equation}

\noindent This differs from previous studies, such as morphological
methods and pair methods, where the merger fraction is defined as the number 
of mergers ($N_{\rm m}$)
divided by the total number of galaxies.  It may appear that these
two are the same quantity, however they are not, as a merger by definition
involves the conglomeration of two or more galaxies. 
Our definition of $f_{\rm gm}$ is
more general, can be easily calculated from the pair definition, and
is directly translated into merger rates using the time-scale
of a merger.  Since usually $N_{\rm gm} = 2 \times N_{\rm m}$, the galaxy 
merger fraction
can be converted into a merger fraction simply by multiplying by a factor
of two for merger fractions computed in 
pair studies.  

The galaxy merger fraction can be directly computed through a morphological 
method using the observed number of mergers ($N_{\rm m}$), and the
total number of observed galaxies ($N_{\rm T}$) by,

\begin{equation}
f_{\rm gm} = \frac{N_{\rm m} \times \kappa}{N_{\rm T} + (\kappa-1)N_{\rm m}},
\end{equation}

\noindent where $\kappa$ is the average number of galaxies that merged
to produce the $N_{\rm m}$ mergers, and must be $\geq 2$.  
We will only use values of $\kappa = 2$ is this paper.  The observational
method for determining $N_{\rm m}$, and the tabulated values, are 
discussed in Conselice et al.
(2000a,b), Bershady et al. (2000), Conselice (2003b)
and Conselice et al. (2003a).

\subsubsection{Merger Fractions at $z < 0.3$}

The pair and merger fraction 
at $z \sim 0$ sets the base-line for understanding
all other higher redshift merger fractions, and it is worth spending some
time trying to determine it, and the corresponding merger rate. When discussing
merger fraction and merger rates, it is imperative to be very specific about
the selection criteria and the luminosities or masses of the galaxies
under study, as the merger fraction changes as a function of mass and 
luminosity in both the nearby and
distant universe (Conselice et al. 2003a; Xu, Sun, He 2004).
 
The traditional method for finding mergers is to search for
systems in pairs, and to quantify this by the pair fraction, $f_{\rm p}$.  
This has been effectively applied up to redshifts 
$z \sim 1$ (e.g., Carlberg et al. 1994; Le F\'{e}vre et al. 2000; 
Patton et al. 2002; Bundy et al. 2004; Lin et al. 2004). 
Using Sloan Digital Sky Survey (SDSS) data, the merger fraction for close 
pairs, with an average galaxy 
luminosity of M$_{g^{*}} = -20.13$,
was computed as 0.5\% (Allam et al. 2004).
A similar merger fraction is found by examining galaxies in
pairs utilizing 2MASS data (Xu et al. 2004).   Other studies, such
as the use of the UGC catalog of nearby galaxies have found that the
merger fraction of galaxies is similar to these values, $f_{\rm m}$ 
= 2.1$\pm0.2$\% (Patton et al. 1997).  

Patton et al. (2000) used a redshift survey of nearby galaxies to 
determine the merger fraction at $z \sim 0$ using new merger quantifiers, 
N$_{c}$
and L$_{c}$, which are the average number of pairs per galaxy (N$_{c}$) and the
average luminosity of these pairs per galaxy (L$_{c}$).  Furthermore, Patton
et al. (2000) show that the vast majority of galaxies with companions are 
pairs, and therefore, $f_{\rm p} = 2 \times N_{c}$.  However, only a fraction 
of these pairs will merge, and
Patton et al. (2000) calculate that at $z \sim 0$ about half of all the
observed pairs will do so.  At high redshifts the fraction likely to 
merger increases as $(1+z)$ (e.g., Le F\'{e}vre et al. 2000).  

On Figure~1, we plot merger fractions, derived from pairs at $z \sim 0$ 
systems taken from Patton et al. (2000) at luminosities M$_{\rm B} < -19$ and
M$_{\rm B} < -21$.  Patton et al. (2002) find that pair fractions, and
the value of
N$_{c}$ increase for fainter systems, and is
effectively zero at bright (M$_{\rm B} < -20$)
magnitude limits.  We convert N$_{c}$ into our 
f$_{\rm gm}$ by multiplying by two to account for the second galaxy involved 
in the merger, and divide by two to account for the relative fraction of pairs
that are likely to merge (Patton et al. 2000).  Thus, in our formalism
$f_{\rm gm}(z = 0)$ = N$_{c} (z = 0)$.

\subsubsection{Merger Fractions from $0.3 < z < 1$}

The merger fraction and rate was first deduced to have been higher in the past
using simple arguments in Toomre (1977). The first measurements of
the merger fraction at high redshift, as opposed to simply inferring the
evolution in the merger fraction (Zepf \& Koo 1989), was performed
by Carlberg, Pritchet \& Infante (1994).  These early merger studies 
were done by finding pairs that were separated
by various angular size, usually 6\arcsec\, or so (Zepf \& Koo 1989).  A major
result from Carlberg et al. (1994) and others (e.g., Patton et al. 1997)
was that the pair fraction is around three times higher at $z \sim 0.4$
than it is today.  These studies characterized the merger fraction
evolution as a power-law with redshifts: $f_{\rm m}$ = f$_{0} \times 
(1+z)^{m}$, with fitted power-law indices, $m$ $\sim$ 3-4.\

Patton et al. (1997) used the now standard definition of pairs as
systems with a separation of 20 $h^{-1}$ kpc, or less, and found
a pair fraction at $z \sim 0.33$ of 4.7$\pm 0.9$\%, while at
$z \sim 0$ the fraction is 2.1$\pm0.2$\%.  In other studies, 
merger fractions were reported to be similar, although
there are some who find a shallower rise in the pair fraction
 up to $z \sim 1$ (Carlberg et al. 2000;
Lin et al. 2005). Much of this discrepancy can be accounted for
by the different luminosity ranges used, and assumptions for how to 
account for luminosity evolution (Patton et al. 2002; Lin
et al. 2005).

Patton et al. (2002) using the formalism for mergers outlined in
Patton et al. (2000) (\S 3.1.2),
computed values of N$_{c} = 0.012\pm0.003$ for galaxies brighter
than M$_{\rm B} = -19$ at $z \sim 0.3$. As at $z \sim 0$, the value of
N$_{c}$ declines at brighter magnitude limits.
Other methods for determining the
pair and merger fraction up to $z \sim 1$ include Le F\'{e}vre et al. (2000)
and Lin et al. (2005), who find similar results. 

The derived galaxy merger fractions found in these pair studies are plotted on 
Figure~1.   The
increase is well fit by a $(1+z)$ power-law, giving indices $m = 2 - 4$,
consistent with most theoretical predictions of how merger fractions 
decline with
redshift (e.g., Gottlober et al. 2001).  However, due to the onset of
the Hubble sequence, it is likely that
most merging activity is over by $z \sim 1$ for massive galaxies 
(Conselice et al. 2005).  Since
50-75\% of the stellar mass in galaxies is formed by $z \sim 1$ (e.g.,
Dickinson et al. 2003), it is
critical to try to place constraints on the formation modes for
galaxies at earlier times.

\subsubsection{Merger Fractions at $z > 1$}

Merger fractions at $z > 1.2$ are currently all determined through 
morphological methods (Conselice et al. 2003a; Lotz et al. 2004).  
The merger fraction is determined
through the use of equation (2),
which includes determining the
observed number of galaxies likely to be mergers through a 
particular computational method.  In this paper, we use the CAS system to 
identify galaxies that are likely
undergoing major mergers at all redshifts (see Conselice 2003b).
Conselice et al. (2003a) presents the detailed reasoning behind
these computations, as well as the $z > 1.5$ data we use in this paper.

All morphological methods identify a number of `morphological 
mergers' ($N_{\rm morph}$) that must be converted into the number of
actually mergers ($N_{\rm m}$) by,

\begin{equation}
N_{\rm m} = N_{\rm morph} \times \frac{f_{1}}{f_{2}},
\end{equation}

\noindent where $f_{1}$ is the fraction of galaxies
identified as a merger that are actual mergers, and $f_{2}$ is the fraction of
actual mergers picked up by the morphological method.  In Conselice et al.
(2003a) it was assumed that $f_{1}/f_{2} = 1$, which for the reasons
below, we use here as well.

The CAS system identifies major mergers through
the use of the asymmetry ($A$) (Conselice 1997; Conselice et al. 2000a)
and clumpiness parameters ($S$) (Conselice 2003).  The
assumptions we make in using these two parameters are that (1) the structures
of galaxies have a physical meaning, (2) star formation occurs in clumps,
as it does in the nearby universe (Lada \& Lada 2003) and 
(3) a galaxy which has a large
scale asymmetry, indicates a non-equilibrium dynamical state, and is likely
undergoing a major merger.  These assumptions are well established
for nearby galaxies (Conselice 2003b), and are likely valid at 
higher redshifts as well.  The quantitative 
criteria used to determine if a galaxy is a merger at $z = 0$ is,

\begin{equation}
(A_{\rm optical} > 0.35) \wedge  (A_{\rm optical} > S_{\rm optical}),
\end{equation}

\noindent where $\wedge$ is the operation definition of `and'. 
Galaxies with $A > 0.35$ in the nearby universe are about 99\%
mergers, and thus f$_{1} \sim 1$ at $z \sim 0$.  The second criteria in 
eq. (4) is important, as galaxies can be asymmetric because of
star formation, as well as from the presence of a merger (Conselice
et al. 2000a).    However, galaxies dominated by star formation, and which 
are not merging, have large clumpiness and asymmetry values because
the light is distributed in localized and compact 
star forming complexes (Conselice 2003b; Mobasher et al. 2004).  
We therefore use the
clumpiness index ($S$) to remove star forming galaxies from consideration as
a merger by the above criteria.  If a galaxy has a high asymmetry, but a low
clumpiness value, it implies that the asymmetric light is not localized,
but is a large scale feature indicating that the system is 
not viralized.  The criteria for this is simply $A > S$.
Not all galaxies involved in a major merger are identified through this
process, yet the simulations discussed in \S 3.3 suggest that all major
mergers are asymmetric at some point in their evolution. Thus, f$_{2} \sim
1$, although it would be lower if we considered mergers within a given 
time-scale.

Merger fractions at $z > 1$ are presented in Conselice et al. (2003a) although
we reevaluate these using equation (2) to obtain the galaxy merger
fraction.  Galaxy merger fractions computed
using two magnitude (M$_{\rm B} < -19$, M$_{\rm B} < -21$) and mass
(M$_{*} > 10^{9}$ \solm, M$_{*} > 10^{10}$ \solm) limits
are shown in Figure~1.  We assume throughout
that $\kappa = 2$ (eq. 2).  The merger fractions computed through the CAS 
system at  $z \sim 1$ are
within 1 $\sigma$ of the merger fractions computed using the pair
fraction method (Conselice et al. 2003a).   
 The most interesting and important feature of these
merger fractions is that they become quite large
at high redshift for the most luminous and most massive galaxies (Conselice
et al. 2003a), with $f_{\rm gm} = 0.5 - 0.7$ at $z \sim 2.5$.  The implication
from this is that the merger rate is very high for these 
systems, and that this might be a way to form modern galaxies.

\subsubsection{Fitting the Merger Fraction}

Figure~1 plots all published galaxy merger fractions available to
date for mergers at two luminosities, 
M$_{\rm B} < -21$ and M$_{\rm B} < -19$, and
two stellar mass limits, M$_{*} > 10^{10}$ \solm and M$_{*} > 10^{9}$ \solm. 
 Note that this is an inhomogeneous data set as the time-scale in which
these fractions are valid can vary by a substantial amount. However
we later show that these time-scales are  similar for
both the pair and the merger method (\S 3.3) and thus can be
compared fairly.

Traditionally, the evolution of
merger fractions with redshift are fit by a power-law of the form $f_{\rm m} = f_{0}
(1+z)^{m}$, where $f_{0}$ is the merger fraction at $z = 0$
and $m$ is the power-law index.  Fits of $m$ using this formalism
have varied from $m = 0 - 4$ up to $z \sim 1$, although most studies
find values that are around $m = 2-4$ (cf. Bundy et al. 2004 and Lin et al.
2004 when examining mergers in the near infrared, and when accounting for
luminosity evolution).  It is worthwhile to examine
whether a power-law history, which is quickly becoming the standard way
to characterize the merger fraction evolution, is in fact the appropriate 
formalism.  

Fitting the merger
fraction by a power-law was initially motivated by the theory of structure
formation.   By assuming
that primordial density perturbations are
Gaussian, the resulting merging history of dark halos
can be understood based on the Press-Schechter (P-S) 
(Press \& Schecter 1974) formalism.
P-S describe how the density of dark matter halos of 
mass $M$ evolve as a function of time.    
The P-S formalism, and its extension 
(Bond et al. 1991; Bower 1991; Lacey \& Cole 1993), agree well with N-body
models of the hierarchical galaxy formation process (e.g., Gottlober et al.
2001), and has been used as the basis for all semi-analytic CDM models,
some of which have published these 
merger histories (Kolatt et al. 2000; Gottlober, 
Klypin \& Kravtsov 2001; Khochfar \& Burkert 2001).

Based on the output of these simulations, the fraction of galaxies
merging is well approximated by the $(1+z)^{m}$ 
formalism up to about $z \sim 2$.   The $m$ index contains much information, 
including the value of the primordial power-law index ($n$), the cosmological
constant, and the matter density parameter ($\Omega_{\rm m}$) 
(Carlberg 1991).
The merger fraction however does not appear to be well fit by a power-law 
empirically out to $z \sim 3$ (Conselice et al. 2003a), especially for 
lower mass and fainter systems.  One fitting function with
some theoretical basis is a mixed 
power-law, exponential function,

\begin{equation}
f_{\rm gm} (z) = \alpha (1+z)^{m} \times {\rm exp}(\beta(1+z)).
\end{equation}

\noindent where the $z = 0$ merger fraction is given by $f_{\rm gm} (z=0) 
= \alpha \times {\rm exp} (\beta)$.  An analytic formulation of the merger
fraction gives a similar form based on the P-S theory 
(Carlberg 1990), and equation (5) is in fact often a better fit than a
$(1+z)$ power-law.

We fit values of $\alpha$, $\beta$ and $m$ from equation (5) for the galaxy
merger fraction in the mass and
luminosity bins used in Conselice et al. (2003a).  We find
$\alpha = 0.2-0.6$, $\beta = -1.5 - -3.7$ and $m = 5-10$ for systems with 
M$_{*} < 10^{10}$ \solm.  
We also fit the traditional $(1+z)^{m}$ fitting formalism, although only 
up to $z \sim 1.5$, or $z \sim 1$ for all but the brightest and most massive
bins.  We find that the merger fraction out to $z \sim 1-2$ can be
well fit by a power-law with power-law indices $m = 2-4$ for all galaxy
types.

There are a few additional points about the merger fraction history that
should be noted.  The first is that for M$_{*} < 10^{10}$ \solm and
M$_{\rm B} > -20$ galaxies there appears to be a peak turnover 
redshift around $z_{\rm turn} \sim 1.5 - 2$ where the galaxy merger fraction 
declines at $z > z_{\rm turn}$ (Figure~1; Conselice et al. 2003a).  
This redshift can be computed analytically through 
d$f_{\rm gm}$/d$z$ = 0, or $z_{\rm turn} = -1 \times (m/\beta + 1)$, where
we compute $z_{\rm turn} = 1.5-2.5$ for galaxies with M$_{*} < 10^{10}$ \solm
and M$_{\rm B} > -21$.  There does not appear
to be a turnover redshift for the most luminous (M$_{\rm B} < -21$) and most
massive (M $> 10^{10}$ \solm) galaxies up to $z \sim 3$, although they
must flatten off at some higher unknown redshift.

\subsection{Merger Rates}

Using the merger fraction  in a physical context
requires that we understand the time scale in which a merger is
occurring, and thus convert the galaxy merger fraction $f_{\rm gm}$ 
into a galaxy merger rate ($\Re$), defined  within a redshift, luminosity, and
mass range.  The formalism for this is,

\begin{equation}
\Re (z, {\rm M_{\rm B}}, M_{\star}) = f_{\rm gm} \tau_{\rm m}^{-1} n_{\rm m},
\end{equation}

\noindent where $\tau_{\rm m}$ is the time scale for a merger
to occur and $n_{m}$ is the co-moving or physical density of all galaxies
within a given luminosity or mass range, and at a given redshift. 
Determining $\tau_{\rm m}$ is critical for utilizing the observed 
galaxy merger fraction, and for determining the importance of galaxy
mergers for forming galaxies.
Since the time-scale of a merger is too long to observe directly, we must
use theoretical arguments and models to determine the time-scale of
major mergers.  The co-moving or physical density $n_{\rm m}$ 
can be determined from
observations of the galaxy luminosity or mass function (Papovich et al.
2001; Shapely et al. 2001) or by directly counting the number of galaxies
within a luminosity or mass bin in the sample under study.  
If the former method
is used, it is critical that the selection method be identical to the method
for finding mergers.
We use two methods, described below, for determining the time-scale
of a merger: dynamical friction arguments and N-body
models of the merging process.

\subsubsection{Dynamical Friction Time-Scales}

The typical method for determining the merger time-scale of two galaxies 
 is the dynamical friction argument. This is based on the
assumption that galaxies are embedded in dark matter halos.  We revisit
this computation here, although a more detailed analysis requires
the use of simulations which we discuss in \S 3.3.

To compute the time required for two galaxies to merge necessitates that
we make some assumptions. The first is that the mass profile of
galaxies are isothermal with a mass distribution that falls
of as r$^{-2}$. Based on this, there is a fictional drag force
that induces angular momentum loss.  As such the
two galaxies will gradually approach each other until they merge.  The
time for two galaxies, separated by $r_{i}$, to be separated
by $r_{f} < r_{i} $ is,

\begin{equation}
t_{\rm fric} = 0.0014\, {\rm Gyr}\, (r_{i}^{2} - r_{f}^{2}) \left(\frac{v_{c}}{100\, {\rm km\, s^{-1}}}\right) \left(\frac{10^{10}\, {\rm M_{\odot}}}{M}\right),
\end{equation}

\noindent where $v_{c}$ is the relative velocity between the two galaxies,
$M$ is the mean accreted mass, and where we have assumed
the Coulomb logarithm, ln $\Lambda$ = 2, based on equal mass merger
simulations (Dubinksi, Mihos \& Hernquist 1999; Patton et al. 2000).
For the velocity ($v_{c}$) and mass quantities ($M$), we take the
average values from Patton et al. (2002), finding $v_{c} = 260$ km s$^{-1}$
and M $\sim 3 \times 10^{10}$ \solm, although these values will
 change when considering galaxies of different masses and luminosities.  
Using these fiducial values, and the projected
radii of a given pair, we calculate the time-scale for a merger to occur,
or rather for the pair's separation to change from $r_{i}$ to $r_{f}$.

We use equation (7) as the time-scale for the merger rate
for galaxies in pairs.  Typically, we find  that the time for a merger to 
occur by equation (7) is 0.5-1 Gyr, 
although the exact value changes under different assumptions.  This is
one reason why pair fraction methods are very difficult to convert into
a physically meaningful quantity.  Throughout this paper we assume 
that the derived merger time-scales for galaxy pairs has an associated
systematic uncertainty of 0.25 Gyr. The structural method, utilizing the
properties of galaxies after they have merged, has the potential to be 
more robust for quantifying the time-scale of a merger, and thus
also the merger rate.

\subsection{N-Body Models of Galaxy Mergers}

The above argument reveals that the merger time scale for
two galaxies of similar mass is roughly 0.5 Gyr.  Our goal in this
section is to understand if we can do better than this estimate
utilizing the structures of galaxies, and the time-scale in which
a galaxy will be identified as a merger structurally.
To determine the time and mass scale sensitivity of galaxy structures 
to mergers we carry out and analyze a series of N-body models of galaxies
undergoing mergers. These models are described in detail in 
e.g., Mihos \& Hernquist (1996), Dubinski et al. (1999) and Mihos (2001).

The models we use are composed of
dark matter and stars.  We do not include the morphological
effects of star formation in these simulations, as modeling this
aspect is very difficult with an infinite number of possibilities, whose
reflection of actual properties is doubtful and is not straightforward to
model. To test the effect
of star formation, we place fake star forming complexes in these simulations 
by hand, which does
increase the measured asymmetry.  The clumpiness index however also
increases such that the effective
asymmetry $\sim (A - S)$ matches the asymmetry before placing these star 
forming complexes in the N-body simulations (\S 3.1.4). 

The techniques and detailed descriptions of these models 
can be found in Hernquist (1993) and Mihos \& Hernquist (1996).
Each of the N-body models are composed of 294,912 luminous (non-dark matter),
and 65,536 dark matter particles, and are modeled using
the TREESPH hierarchical tree (Hernquist 1987).  Most of our models are
composed of bulges and disks, with a bulge to disk 
ratio of 1:3, while others are pure disks.    
The dark matter halo in these simulations are isothermal
spheres with core radii equal to the scale length of each model's disk. 
The dark
matter halo is truncated at radii larger than 10 times the core radius, with
an exponential decline (Hernquist 1993).  The mass of the disk in these
simulations is M$_{d} = 1$, with a bulge mass of M$_{b} = 1/3$ and a halo
mass of M$_{\rm h} = 5.8$, all in simulation units. Scaling to the Milky Way 
this gives a total mass of 3.2 $\times 10^{11}$ \solm,  and a model 
disk scale-length of 3.5 kpc.  
The ratio of the total masses for the simulated galaxy pairs
in the merger simulations are 1:1, 1:2, 1:3 and 1:5.  
We analyze images of these simulations using snap shots separated by
26 Myrs.
For each simulation we analyze a total of 61 snap shots 
from the beginning of the simulation until $\sim 1.5$ Gyr.  We analyze
ten different simulations (listed in Table~1) at seven different viewing
angles each.

We study these N-body simulations using the same 
structural techniques used to study galaxies at high and low redshift using
the CAS morphological analyses techniques. 
For three mass ratios (1:1, 1:2, 1:3) we simulate
mergers with a mix of the following orbital properties: inclined, retrograde
and prograde. The prograde simulation contains one disk inclined by
20\deg\, to the orbital plane, while the inclined simulation has a 75\deg\,
inclination, and the retrograde is at a 135\deg\, inclination.  
The three simulations at each mass include: one galaxy
which is inclined and another which is retrograde (IR), one galaxy which is 
prograde and another that is inclined
(PI), and simulations with one galaxy prograde and the other retrograde (PR).
Figure~2 shows a graphical representation of the 1:1 IR simulation.

We carry out CAS measurements on these simulations in three different
ways, each designed to match how these values are measured on actual
galaxies.  Different measurement methods are
important to consider as the resulting values of the CAS parameters can
vary depending on how the galaxies were detected and 
cataloged.    The three main issues are: (1) when galaxies are
merging are they detected in the same SExtractor 
segmentation area (Conselice et al.
2003a, 2004, 2005)?, (2) when measuring the CAS parameters during the merger, 
do you remove
the second galaxy? and (3) where the center is placed can change the outcome
of the CAS measurements.

All of these points deserve discussion, although they are inherently related
to each other. The process for measuring the CAS parameters for a galaxy
involves two steps, and because the next generation of morphological analyses
will involve at least tens of thousands of galaxies, this process must be automated.
We have therefore considered various approaches for measuring the CAS
parameters on these simulations.  The detection process on Hubble
Space Telescope imaging almost universally involves the SExtractor program
which detects and splits galaxies using significance thresholds.  Usually
the way this is fine-tuned is to separate galaxies by trial and error using
the detection and separation parameters.  However, it is inevitable that
close galaxies will sometimes be placed in the same detection.  The
prevalence of this needs to be determined by  examining 
galaxies by eye that are asymmetric, although details of this process
are beyond the scope of the current paper.  

However, there are situations were two galaxies are close enough during
a merger that they should be considered the same system.  In Conselice (2003b)
it is argued that the extent of a galaxy should be defined as the light
within the Petrosian radius. If there are two interacting galaxies, which
are spatially close,
then wherever the Petrosian radius converges determines whether or not those
two galaxies should be considered the same system or not. In this situation
the center of the system will be between the two centers of the merging
galaxies.  As such, we also consider this center when performing CAS
measurements, determining the asymmetry time-scale for each scenario. 
 
We list the time-scale information for simulations at a
representative viewing angle of 54\deg, the results of which are listed
in Table~1 as $\tau_{1} - \tau_{4}$.  This is the time in which the simulated
galaxies are asymmetric enough to be considered a major merger using the
same criteria used on high redshift galaxies.
Another scenario we consider is when the center
of the CAS run is placed on the brightest galaxy's center and when the
fainter galaxy or pair is not removed. The time in which this pair will
be seen as an asymmetric merger, with $A > 0.35$, 
is listed in Table~1 as $\tau_{2}$.

Very often however, what occurs in the SExtractor process is
that two galaxies are separated into different systems.  The CAS code
has an option, which is now being used (e.g., Conselice et al. 2004), 
to remove all galaxies
from an image not currently being analyzed. We are then left with a single
galaxy.  To mimic this method we only image one galaxy at
a time in our simulation, and measure its CAS parameters.  This is effectively
the same approach used in Conselice et al. (2003a) where the CAS values
were measured within the area given by the segmentation maps. 
Neighboring galaxies were however not removed, similar
to the sim2 simulations whose time-scales we use in \S 3.3.4 to obtain
merger rates.     One result of the `galaxy removal' simulations
is shown in Figure~3 where the asymmetry, concentration, and
size measurements for the brightest galaxy in a merging pair are shown
as a function of time in the 1:1 IR simulation (Figure~2).
To determine the effects of viewing angle on these time scales,
we viewed the merger simulations at viewing angles of 
15\deg, 54\deg, 77\deg, 90\deg, 117\deg, 145\deg, and 165\deg, finding
the same result to within 0.2 Gyr.

There are several observations to take away from Figure~3.
The first is that for the 1:1, 1:2 and 1:3 simulations each galaxy starts
with a low asymmetry and a high
concentration.  In the 1:1 simulation, as the two galaxies start to 
interact, at about 0.4 Gyr 
(see Figure~2), the structures become more asymmetric.
For an interval of $\sim 0.2$ Gyr these galaxies remain asymmetric enough 
due to this encounter to be identified as an asymmetric merging galaxy.  After
the initial encounter, the galaxies separate and contain a lower
asymmetry until 1 Gyr into the simulation when both galaxies become
asymmetric for another 0.2-0.3 Gyr.  At about 1.3 Gyr the two systems
have completely merged and the merger remnant has cooled enough dynamically to
have a low structural asymmetry.  Note that from Figure~3 the asymmetry
evolution of the 1:1 simulation is to first order independent of orbital
type (Table~1).  Some variation does exist, for
example the 1:1 IR simulation has a delayed asymmetry
response during the second peak, although this does not change the derived
effective merger time-scale.

The CAS merger time-scale sensitivity for the more massive galaxy in this 
configuration ($\tau_{3}$), with the lower mass galaxy not imaged, is tabulated
in Table~1.  The listed values of $\tau_{4}$ are the CAS merger time-scales
for the lower mass of the 
pair, when the more massive system is not imaged. 
We find that the two times are nearly identical, that is 
the time scale is independent of mass, and mass ratio up to 1:3.
The average and 1 $\sigma$ range in the time-intervals 
in which the asymmetry index is sensitive to finding a merger 
are $\tau_{1} = 0.82\pm0.12$, $\tau_{2} = 0.43\pm0.08$, 
$\tau_{3} = 0.29\pm0.20$, and $\tau_{4} = 0.29\pm0.19$. The 1 $\sigma$
systematic range quoted for these different time intervals are taken from the
values in Table~1.   They thus include ranges in mass ratio and orbital
properties, but do not include systematics from different viewing angles,
which adds another 0.1 Gyr uncertainty.  Note that
these 1 $\sigma$ systematic ranges are not proper 1 $\sigma$
error measurements on the time-scales in which galaxies at high redshift
will merge within.  They are only the 1 $\sigma$ range in the merger
time-scales for the galaxy simulations that we study, which likely does
not represent the real range in these galaxies.  

\subsubsection{Pure Disk Simulations}

The simulations discussed in \S 3.3 are for galaxies with a 1:3 bulge to disk
ratio.  The  merger time-scale can vary if the
structure of galaxies are different from the bulge/disk ratios we
assumed for our initial galaxies.
As such, we perform a 1:1 simulation of two pure disk galaxies, the results of
which are shown in Table~1 as 1:1NoB.  We find that the pure disk galaxy
simulations are asymmetric enough to be counted as a merger for a longer
period of time than galaxies with bulges, by as much as a factor of two.  
This is the case
when we consider the simulations where only one galaxy is considered 
($\tau_{3}$
and $\tau_{4}$) as well as when both galaxies are used in the CAS 
computations ($\tau_{1}$
and $\tau_{2}$).   The reason for this is that these pure disks do not
decrease in asymmetry after the first encounter.   
The presence of a bulge is a stabilizing entity that
creates a lower asymmetry when the two simulated
galaxies are in the process
of merging.  This creates some uncertainty in the time-scale for
the merger, although as discussed in \S 3.3.3, this effect is not likely
to be large when we
consider the numerous mergers a massive galaxy undergoes, and the
likely formation of bulge-like systems in massive galaxies by 
$z \sim 3$ (Baugh, Cole \& Frenk 1996; Ferguson et al. 2004).

\subsubsection{Minor Mergers Simulations}

We investigate a minor merger model\footnote{Observationally, a minor
merger is typically a merger which has a mass ratio greater than 1:3 or 1:4,
although dynamicists usually consider a galaxy with a 1:5 ratio an intermediate
mass merger, and only mergers with mass ratios of 1:10 or greater, a
minor merger.}, a 1:5 mass ratio merger, whose resulting asymmetry,
concentration, and size as a function of time is shown in Figure~4. The most
obvious result of this simulation is that neither galaxy ever obtains an 
asymmetry value high
enough to be considered a major merger. The two galaxies do not
effectively merge until 3.5 Gyr into the simulation when the asymmetry
becomes high, but just shy of the $A = 0.35$ limit for a merger at $z = 0$. 
 The time-scale for
this minor merger to complete is much longer than the lower mass ratio major 
mergers, 
with a total merger time of $> 4$ Gyr.  Since the age of the universe 
at $z > 2$  is
small (3.3 Gyr), it is unlikely that a single minor merger can produce
the high asymmetry signal seen in high redshift galaxies.  Although
we caution that more extensive simulations of lower mass ratio mergers with
different initial conditions are
needed to fully determine the generality of this.  This simulation
also likely does not match the initial conditions of
galaxies in pairs, and the distributions of orbital energies for high$-z$
pairs which is currently unknown.  We also cannot
rule out the possibility that multiple minor mergers are occurring at once
to produce a large asymmetry signal. More N-body models of different
merger scenarios are necessary to address these issues.

From the analysis in this section we conclude that the mergers of 
galaxies with mass ratios higher than 1:5 are not likely producing a 
merger signal through the asymmetry index.  If this is indeed the case,
and we assume it is for the remainder of this paper, then it implies
that the CAS method for finding mergers is a powerful approach that
is insensitive to minor mergers, even a 20\% mass merger.   This is
further confirmed empirically through observations of galaxies in
pairs in the local universe (Hernandez-Toledo et al. 2005).

\subsubsection{Limits on the N-body Models}

The simulations we examine above are necessarily limited, and only offer a 
subset of the possible variety of conditions experienced by galaxies undergoing
mergers/interactions at high redshift.  It is currently
not feasible to perform a more general analysis than the above, 
as a full suite of self-consistent N-body simulations with all possible
initial conditions does not yet exit.  The simulations we examine
are low speed encounters with isothermal dark matter profiles.  However, we 
consider several other possibilities, including high-speed encounters
between galaxies, galaxies with dark matter profiles that
differ from isothermal, and galaxies mergers that have different masses than
the nominal Milky Way mass considered here.

First, high speed encounters between galaxies has
been modeled by Moore, Lake \& Katz (1998) through `galaxy harassment'.
These interactions can result in significant mass loss (Conselice 2002),
and is likely ongoing in clusters today (Conselice \& Gallagher 1999; 
Gregg \& West 2004; Adami et al. 2005).  This
may also be an important process for forming dwarf galaxies.  
However, these encounters do not produce a significant
morphological disturbance in a galaxy.  Over time harassment can
destroy disks, but this process has a time scale much longer than the
life-times of the morphological
disturbances we measure with the asymmetry index.  These high-speed
impulsive encounters increase the internal energy of the interacting systems,
which results in a galaxy which physically expands, but does so in a uniform
manner.
We therefore consider it an unlikely possibility that very high speed galaxy 
encounters can produce a large asymmetry signal that we would identify
as a major merger.

Another complication is that galaxy interactions which do not
necessarily merge, can potentially create a large asymmetry that
would be miscounted as a merger.  These false-positive mergers
are unlikely, based on several arguments.  The first is that
empirically only nearby galaxies that are within a radii of each other
produce a large asymmetry signal (Hernandez-Toledo et al. 2005).  As
these nearby pairs are within a few 10s of kpc of each other, and have very
small velocity differences, they are in the processes of merging, as they
simply do not have a high enough velocity to escape.  Galaxies which
are interacting, but are not as close as a scale-length,
do not have asymmetry values in the merger regime (Hernandez-Toledo et al. 
2005).  These results
are also suggested by our N-body models where a similar pattern is seen.
Furthermore,  if  high redshift galaxies
have an initial velocity difference which is large, and therefore repeated
`fly-bys' occur until a final merger takes place, only the low
velocity encounters will produce a large asymmetry. Based on our models
we include only one fly-by.  However,
we consider in the analysis in \S 3.3.4 the asymmetry merger
time-scale becoming larger due to more than one fly-by.

A further complication in these models is that the mass of 
the most massive system is similar to the Milky Way.  The dynamical
friction time scale for two galaxies to merge is largely 
independent of the mass ratio
of the two galaxies, as long as they can be considered point-like
particles.  However, once the two systems begin to interact strongly
the time-scale can vary (Hernandez \& Lee 2004).  
After the two galaxies merge, the relaxation
time-scale ($\tau_{\rm relax}$) 
depends on the density of the merger remnant such
that $$\tau_{\rm relax} = ({\rm R^{3}}/{\rm GM_{\rm dyn}}) \,\alpha\, 1/({\rm G \rho}).$$  Using
the Fundamental Plane relation we know that the luminosity of a 
galaxy is proportional to a factor ($\beta$) of its velocity dispersion, 
or $${\rm L}\, \alpha\, {\sigma}^{\beta}.$$  N-body models
suggest that merger remnants follow this relationship (e.g., Capelato et al.
1995; Aceves \& 
Velazquez 2004; Boylan-Kolchin, Ma \& Quataert 2005).  Using
the relationship $\rm M_{dyn} \,\alpha\, \sigma^{2}{R}$ and
the fact that ${\rm L} = ({\rm L/M_{\rm dyn}}) \times {\rm M_{\rm dyn}}$, we
can write the ratio of the relaxation time-scale for merger remnants at
two different masses as,

\begin{equation}
\frac{\tau_{\rm relax}}{\tau'_{relax}} = \left( \frac{\rm M_{dyn}}{\rm M'_{dyn}}\right)^{1-(3/\beta)} = \left( \frac{\rm M_{dyn}}{\rm M'_{dyn}}\right)^{1/4},
\end{equation}

\noindent assuming that the ratio $({\rm L/M_{\rm dyn}})$ is independent of
M$_{\rm dyn}$.  By using the value $\beta = 4$ found through the Sloan Digital
Sky Survey (Bernardi et al. 2003), we obtain the second part of equation (8),
or that the time-scale increases as M$_{\rm dyn}^{1/4}$.  This suggests
that the asymmetry-time scale during relaxation from a merger has a
weak dependence on the total mass.  We account for this dependence in \S 3.3.4.

Finally, it is possible that the dark matter halos of galaxies
at high redshift are different than the ones used in our simulations. 
One possibility is that the dark matter profile is less
step and dense than an isothermal profile, such that the dynamical friction
time-scale is shorter.  Using a variety of reasonable 
possible dark matter profiles, Mihos, Dubinksi \& Hernquist (1998) and
Dubinski, Mihos \& Hernquist (1999) investigate how different profiles
will change the structure of tidal tails and features in merger galaxies.
These papers, particular Dubinski et al. (1999), show that extended profiles
do not produce tidal tails, however the inner parts of these galaxies
are distorted due to the merger, irrelevant of the dark matter profile,
and this is the part of the galaxy that contributes the most to the
asymmetry signal.  We have not analyzed these simulations since the parameter
space of possible dark matter halo profiles at high redshift is infinite, as
there are currently no observational constraints. It therefore remains an 
unaccounted for systematic in our merger time-scales.

\subsubsection{Computed Merger Time-Scales and Rates}

We can use the above information to determine a few very important quantities
that have up until now been largely unknown.  The first is the time
in which two galaxies merging will be identified as a merger within the CAS
system.   In Section 3.3, we describe the
total amount of time our simulated galaxies appear asymmetric, yet we
can divide this time further into the time a galaxy is asymmetric
due to a `fly-by' encounter, and before the galaxy relaxes after the merger 
occurs.    We therefore divide the CAS merger time-scale 
into two different components: a fly-by
asymmetry time ($\tau_{\rm fly,A}$), and an asymmetry relaxation time 
($\tau_{\rm relax,A}$). Note that these two quantities are solely define 
as the time when the asymmetry value ($A$) is $>$ 0.35.  In the simulations
we analyze, the merger begins to relax after 0.7 Gyr, which we use as
the dividing line between the fly-by and relaxation time-scales. 
If there is more than one fly-by before the final
merger, then the time-scale will include several factors of 
$\tau_{\rm fly,A}$.  We denote the number of fly-by interactions as 
$N_{\rm fly}$ and express the merger asymmetry time scale $(\tau_{\rm merger,A})$as:

\begin{equation}
\tau_{\rm merger,A} = ({N_{\rm fly}}) (\tau_{\rm fly,A}) + (\tau_{\rm relax,A})
\end{equation}

\noindent To calculate merger rates using $\tau_{\rm merger,A}$ 
we will use the average results of the major merger simulations
where the pair of the galaxy under studied was not removed, namely $\tau_{2}$ 
(Table~1) with a total asymmetry merger time-scale of $0.43\pm0.08$.  This is
the simulation where the analysis mode matches the method used to derive the 
merger fractions in the Hubble Deep Fields.
We find on average that the fly-by time scale is  $\tau_{\rm fly,A}$
$\sim$ 0.23$\pm$0.05 Gyr, which we find is relatively independent of mass 
ratio.
Because this time is due to dynamical friction, it is also 
relatively independent of
total mass (Carlberg et al. 2000).  We further find from the $\tau_{2}$ 
simulations that
the average relaxation time-scale is $\tau_{\rm relax,A}$
= 0.20$\pm$0.05 Gyr, which is also 
largely independent of the merger
mass ratio.  The small error range on these numbers results
from using sim2 from Table~1, whose total asymmetry time range is 0.08 Gyr
(\S 3.3).   This is likely an underestimate, as the actual range in physical
properties is certainly larger than what current N-body simulations 
can provide.  Later, we add uncertainties from the merger fractions, and
the 0.1 Gyr uncertainty due to the viewing angle, to these errors when 
computing merger rates and the merger history (\S 4).

From equation (8), the value of $\tau_{\rm relax,A}$ also
depends slightly on the mass of the merger remnant.  As the total mass of our
simulated galaxies are 3.25 $\times 10^{11}$ \solm, we can write the asymmetry
merger time scale as a function of mass (${\rm M_{tot}}$) as, 

$$\tau_{\rm m} = \tau_{\rm merger,A} = (0.23\pm0.05)({N_{\rm fly}}) +$$ 
\begin{equation}
\hspace{3cm} (0.15\pm0.05)\left(\frac{\rm M_{tot}}{10^{11} {\rm M_{\odot}}}\right)^{1/4}.
\end{equation}

\noindent We use equation (10) for the asymmetry merger time-scale in the 
following calculations. 
For the reasons discussed in \S 3.3.3 we will only
consider cases when $N_{\rm fly} = 1$. 
There are two other systematic errors that we also
consider. First, there is a possible systematic increase
in the merger time-scale of $+0.35$ Gyr if the galaxies undergoing mergers do
not include a bulge.  Observations suggest that the most massive galaxies
at $z \sim 2.5$ have a central light concentration, often consistent with
bulge-like features (Conselice et al. 2005; Ravindranath et al. submitted),
thus it is not likely that many of the massive galaxies  merging at this time
are pure disks.
Simulations also suggest that modern ellipticals  formed from
bulge like systems, as pure disk mergers do not produce correct 
elliptical galaxy scaling relationships (Gonzalez-Garcia \& Balcells 2005).
The time-scale given by equation (10) is  also
similar to that found empirically for real disk-disk mergers
(Hernandez-Toledo et al. 2005).

We can then derive from equation (10) the rate of
galaxy merging for systems with different masses and luminosities (Figure~5). 
To calculate merger rates we use eq. (6) and eq. (10), where for a Milky Way
mass galaxy, $\tau_{\rm m} = 0.43$ Gyr.  We use in equation (6) the best-fit
merger fraction values as a function of redshift, $f_{\rm gm}(z)$, as 
discussed in \S 3.1.   We assume $M_{\rm tot}/M_{*} = 10$ and $M_{*}/L_{\rm B}
= 5$ to convert luminosities and stellar masses into total masses. The
number densities of galaxies at various redshifts used in eq. (6)
are computed using the observed total number of galaxies
in these various calculations (Conselice et al. 2003a,c; Patton et al. 2000;
Patton et al. 2002).  Finally, Figure~5 shows the calculated merger rates in
co-moving volume units and in physical volume units.

Figure~5 shows that the merger rate at all
luminosities and masses  is relatively constant from $z \sim 3$
to $z \sim 1$, but drops rapidly at $z < 1$.  This implies that from
$z \sim 3$ to $z \sim 1$ galaxy mergers are very common, but at later times 
they drop quickly.  This can be seen directly in the galaxy population.  
The gross
morphologies of galaxies at $z \sim 1$ are very similar to their
distributions and properties at $z \sim 0$ (Conselice et al. 2005).  
This would not be the
case if major mergers were still occurring in large numbers, as spiral
galaxies would rapidly
evolve morphologically as they merge (e.g., Hernandez-Toledo et al. 2005).

Furthermore, Figure~5 shows that the merger rate is high for lower-mass
and fainter galaxies, such that there are more major mergers occurring per
unit time, per unit volume, than for the most massive systems. However,
because there are so many low-mass and faint galaxies at all redshifts,
the merger fraction is lower than for massive and luminous
galaxies (Figure~1; Conselice et al. 2003b). An important
question to ask is why the merger fraction is so much lower for the
lower mass and less luminous galaxies. One possibility is that because
higher mass and more luminous galaxies cluster more strongly than the 
lower mass and lower luminosity systems (e.g., Giavalisco \& Dickinson 2001;
Adelberger et al. 2005) the most massive systems are more likely to
merge.  Minor mergers are likely playing
some role in the formation of galaxies at the redshifts we study, however
we cannot place constraints on this process using the current techniques.

Another result from our simulations is that the CAS methodology for
finding mergers can be successfully used to find major mergers.   This means 
we can make reasonable assumptions about how much mass
is added to galaxies identified as ongoing mergers.  We hereafter
assume that the average accreted galaxy pair is 65\% of the mass of
the original, or that the mass ratio of mergers are 1:1.5 in the calculations
that follow.

\section{Massive Galaxy Evolution Due to Mergers}

\subsection{The Major Merger History of Massive Galaxies}

Understanding the modes of star formation at high redshift, that is
determining what triggers the formation of stars, is still largely 
uncertain.  One of
the first attempts to quantify this was in Conselice et al. (2003a, 2005) where
it was argued that a significant fraction of galaxies at high
redshift are undergoing mergers.  We further argued that the merger
rate and mass accretion rate due to mergers can be computed using
assumptions about the merger time-scale at high redshift.  Thus, using the
results from the N-body simulations discussed above, we can now
tentatively calculate the history of galaxy merging for the first time.

If we assume that $z < 3$ galaxies are undergoing major mergers in
the quantitative way described
earlier through the computed galaxy merger fractions, then we can use
this characterization to determine the average number of major mergers 
an average galaxy of a given mass at $z \sim 3$ will undergo
by the time it reaches $z \sim 0$.  This is computed by integrating
the merger rate divided by the density; or
the fraction of galaxies observed at each redshift undergoing
a major merger divided by the time scale in which a merger remains 
identifiable as a merger ($\tau_{\rm m}$).    
By integrating this, we obtain the number
of mergers an average galaxy undergoes between $z_1$ and $z_2$ ($N_{\rm m}$),

\begin{equation}
N_{\rm m} = \int^{z_2}_{z_1} \frac{f_{\rm gm}(z)}{\tau_{\rm m}} {\rm dt} = \int^{z_2}_{z_1} t_{H} \left(\frac{f_{0}}{\tau_{\rm m}}\right) (1+z)^{\rm m_{A}-1} \frac{\rm dz}{{\rm E}(z)},
\end{equation}

\noindent where $f_{\rm gm}$ is the galaxy merger fraction, $\tau_{\rm m}$ 
is the asymmetry merger
time scale (eq. 10), and the parameter E($z$) = ($\Omega_{M}(1+z)^3 + \Omega_{k}(1+z)^2 
+ \Omega_{\Lambda})^{-1/2}$ = H$^{-1}$($z$). We have assumed a power-law 
increase for the form of $f_{\rm gm}$, since this is the best fitting 
formula for the most massive and most 
luminous galaxies.  This is easy to change when considering  
power-law/exponential fits for lower mass galaxies, which we do when 
determining the number of mergers and the resulting
mass accreted in these systems.  This calculation also requires that we
track the evolution of the stellar masses of these galaxies, as the values
of m$_{\rm A}$ and $f_{0}$ evolve with time and stellar mass. However, for
the most massive galaxies we do not have the mass resolution at M$_{\star}
> 10^{10}$ \solm to carry out this solution.  Future studies that utilize 
larger samples will have
this resolution, and a formal solution utilizing eq. (11) can be applied
with the evolving merger history included.

According to this formalism, and using the best fit values for
m$_{\rm A}$ and $f_{0}$, we calculate the merger history of galaxies
at various initial masses and luminosities.  Figure~6 
shows the cumulative number of mergers an average
galaxy with initial stellar masses of 10$^{10}$ \solm, 10$^{9}$ \solm, and
10$^{8}$ \solm,  undergoes as a function of redshift, starting
at $z \sim 3$.   Just as in \S 3.3.4, we utilize the merger time-scale
from sim2 (Table~1) and include in the error budget the time-scale
range from different orbital properties, mass ratios and viewing angles.

Based on this, we find that a 10$^{10}$ \solm galaxy will experience 4.4$^{+1.6}_{-0.9}$ 
major mergers according to this formalism by $z \sim 0$ using equations
(10-11).  If we consider that mergers occur with two fly-bys instead of the
one we assume, then the total number of mergers will be $\sim$ 3, and if
we assume that all of these mergers occur with pure disks as progenitors, 
then an average system will undergo $\sim 3$ mergers.  However, as argued in 
\S 3.3.3 and above, neither of these extreme scenarios are likely.  

It is possible that the merger time-scales for galaxies
are shorter at higher redshifts than at $z \sim 0$, which would imply that the
asymmetry time-scale $\tau_{\rm m}$ would also be shorter.  We use the
observed evolution of the sizes of high redshift galaxies to approximate
how the merger time-scale could possibly change as a function of redshift.
Ferguson et al. (2004) find that the sizes of Lyman-break galaxies
evolve with redshift such that the average measured radius increases
as H$^{-1}$($z$) = E($z$) = (0.3(1+$z$)$^{3}$ + 0.7)$^{-1/2}$ in our
cosmology.  If the merger time-scales increases linearly with radius, then
$\tau_{\rm m}(z)$ $\sim$ E($z$).  When we consider this evolution of
$\tau$ with redshift we find that
the most massive galaxies undergo more mergers between $z = 3$ and $z = 0$, 
roughly 15.2$^{+5.4}_{-3.2}$.   We also
calculate these curves for the lower mass galaxies using the
power-law/exponential formalism for the merger fraction history $f_{\rm gm}$
in eq. (11).  This is however an extreme scenario and there is no evidence
that the merger time-scale at high redshift is any shorter than it is at
$z \sim 0$.

Interestingly, in all cases the lower mass galaxies experience a similar 
number of mergers by $z \sim 0$ (Figure~6).
The major difference is that for the most massive galaxies, most of
these mergers occur earlier at $z > 1$, while the lower mass systems
have few major mergers at similar redshifts.  Most merging for
galaxies with M $> 10^{10}$ \solm appears complete by $z \sim 1.5$, 
with no major mergers after this time.  This implies that massive galaxy
formation is complete by $z \sim 1.5$ if major mergers are the dominate
mechanism for forming high mass systems.

\subsection{Stellar Mass Evolution of Field Galaxies}

The stellar mass of a galaxy undergoing a merger increases due
to the stellar mass accreted in the merger, as well as from any star formation
induced during the merger. Below we present a very general calculation 
and formalism for calculating
 the way stellar mass can be increased in galaxies during the merger process.
Since galaxies at high redshift are
thought to be gas rich systems, the amount of stars produced during
a burst can be quite significant.  There are some estimates for
the ongoing and past star formation rates in Lyman-Break galaxies that
we use to determine how much stellar mass is added due to the star formation
induced during a merger.
In general, the total amount of stellar mass added to a galaxy over
time is given by $\delta M_{\rm T}$, $$\delta M_{\rm T} = 
\delta M_{\rm msf} + \delta M_{\rm asf} + \delta M_{\rm merger}.$$

\noindent where $\delta$M$_{\rm msf}$ and $\delta$M$_{\rm asf}$ are the 
amounts of stellar mass
added due to star formation induced by the merger and from gas accretion, 
respectively, while $\delta$M$_{\rm merger}$ is the amount of stellar
mass added due
to the merger process.  Since from \S 3.3 we know that only mergers
with mass ratios of 1:3 or less will produce the  signal for a major
merger in the CAS system, the amount of mass added in a merger
detected through the CAS method must be similar to the original galaxy's 
mass.  

The amount of stellar mass added due to star formation induced from a merger
can be calculated from fitting the spectral energy distributions (SEDs) of high
redshifts star forming galaxies to model star formation histories.  By
doing this, the best form for the star formation history can be retrieved.
Several studies have investigated the stellar populations, star formation
history, and stellar masses of galaxies at $z \sim 2-3$.  What is generally
found is that the star formation history can be fit as either constant,
exponential, or in bursts, but with the exponential model the most
general and best fit (Papovich et al. 2001). 

In Papovich et al. (2001) and Shapley et al. (2001), the spectral energy 
distributions of LBGs are fit to a star formation model that gives the 
current star formation rate ($\Psi$), the e-folding time of the starburst 
($\tau_{\rm sf}$), and the
current stellar mass at solar and 0.2 solar metallicities, and with Salpeter
and Scalo initial mass functions (Shapley et al. (2001) only fit for
solar metallicity however).  These models generally assume that
the star formation history exponentially declines from an initial star 
formation rate $\Psi_{0}$,
such that the star formation rate at a given later time $t$ is,

\begin{equation}
\Psi(t) = \Psi_{0} \times {\rm exp} (-t/\tau_{\rm sf}).
\end{equation}

\noindent This fitting also gives an age for the burst (t$_{\rm sf}$).  
This is however
the star formation law for the most recent burst of star formation.  If
there are several episodes of star formation this is not
easily revealed through the SEDs of galaxies.
One problem with these fits is that they are based only on SEDs out to
the observed K-band, and thus could in principle be missing a significant
amount of older stellar mass. Preliminary results using {\it Spitzer}
observations reveal that there is not a large missing old stellar
population in Lyman-break galaxies, and the fitted $\Psi_{0}$ and
$\tau_{\rm sf}$ parameters do not differ significantly after
adding in the rest-frame
near infrared to the SED fitting (Barmby et al. 2004).

We perform calculations of how much stellar mass is added to galaxies
due to induced star formation with some trepidation
since the exact form of the star formation history of galaxies is still
largely undetermined from observations, and may not have a unique solution,
or standard form. The following results should
be taken as a general first step at solving this problem, and not as a final
quantitative result. 

Although the starbursts we see
in LBGs are ongoing, and thus have already created some of their stellar
material, we use these fits to predict how much stellar mass will be
created through future events by assuming future star formation will
be similar, and induced during each merger.  The amount of stellar mass created
through starbursts induced through mergers ($\delta$M$_{\rm msf}$) is then,

$${\rm \delta M_{msf}} = \int^{z_2}_{z_1} \int^{t_{\rm m}}_{0}  t_{H} \left(\frac{f_{0}}{\tau_{\rm m}}\right) (1+z)^{\rm m_{A}-1} \frac{\rm dz}{{\rm E}(z)} 
\times$$
\begin{equation}
\hspace{2cm} \Psi_{0} {\rm exp}(-t/\tau_{\rm sf}) {\rm dt},
\end{equation}

\noindent where $z_2$ and $z_1$ are the final and observed redshifts,
respectively, $t_{H}$ is the Hubble time,  and $t_{\rm m}$ is
the time from the onset of the merger until the present day, and for
integration purposes can be effectively considered infinite.  When
$t_{\rm m} >> \tau_{\rm sf}$ we can approximate the star formation part of 
equation (13) as $\Psi_{0} \tau_{\rm sf}$.  We can then compute the 
amount of stellar mass 
added during mergers due to induced star formation
as $\delta$M$_{\rm msf} = \Psi_{0} \tau_{\rm sf} N_{\rm m}$, where 
$N_{\rm m}$ is the number of major mergers.  For these calculations we use 
the median
values, $\tau_{\rm msf}$ = 20 Myrs, and an ongoing star formation
rate $\Psi_{0}$ = 32 M$_{0}$ yr$^{-1}$, for $z > 2$ LBGs
(Papovich et al. 2001). 
 We normalized the initial star formation
rate by computing the average specific star formation rate for the Papovich
et al. sample and calculated the actual star formation rate for each 
galaxy mass by assuming that the specific star formation rate
is scale free. We obtain very similar results if we use the average star
formation rate for all stellar masses.  We also include the amount of stellar 
mass
added from the ongoing merger ($\delta M_{\rm asf}$) observed in
each system through a similar method.  There are some caveats to this
approach. The first is that the values of $\tau_{\rm msf}$ in Papovich
et al. are not well constrained.  As such, it is possible that our 
reconstructed
star formation histories are not representative of the actual stellar
mass produced in high redshift star formation events. An alternative method
is to assume that the star formation is produced in bursts within a given
duration at a constant rate.  The implementation of this
formalism results in effectively the same mass formed as the exponential star 
formation rate over an infinite (or very long) period of time.  If the burst
duration is 100 Myr, then the amount of mass formed would be roughly
three times higher than the exponential star formation decline.

To obtain the stellar mass added to a given galaxy through the mergers
themselves, we tried two method, which are nearly identical. First, we used the
empirical formalism in Conselice et al. (2003a) which gives the amount of 
stellar mass accretion onto a galaxy due to major mergers as a function of 
redshift.  This can be generalized as:

\begin{equation}
{\rm \delta M_{merger}} = \int^{z_2}_{z_1} \frac{1}{f_{0}} (1+z)^{m_{A}} \frac{\Delta {\rm M}}{\tau_{\rm m}} (1+z)^{m_{\rm M}} {\rm dt}
\end{equation}

\noindent where $\rm {dt}$, as before, is given by ${\rm dt = t_{H}} \times 
dz/[(1+z){\rm E}(z)]$ and
$\Delta {\rm M} (1+z)^{m_{\rm M}}$ is the empirically calculated stellar 
mass added to a galaxy due to major mergers every Gyr per galaxy.  We 
also estimated the number of mergers a galaxy undergoes using
the relationship $\delta M_{\rm merger} \sim 2^{N_{\rm m}} \times {\rm M}_{0}$
for equal mass mergers, or in our case $1.65^{N_{\rm m}} \times {\rm M}_{0}$ 
for 1:1.5 
mass ratio mergers, where M$_{0}$ is the initial stellar mass of the galaxy.
Both methods give similar stellar mass accretion rates from the major
merger process. 

By using $\tau_{\rm m}$ from equation (10), and the values 
$\rm m_{A}$ = 2.7$\pm$0.5, $f_{0} = 0.01$, and $m_{\rm m}$ = 1.47, based on 
our empirical fits discussed in this paper for galaxies with 
M$_{*} > 10^{10}$ \solm, we compute the final $z \sim 0$ stellar mass of an 
average massive (M$_{*} > 10^{10}$ \solm) $z = 3$ galaxy.  The
result of these calculations is show in Figure~7 where the stellar mass
build up for galaxies of different initial stellar masses are shown.  Based on
this, it can be seen that the most massive galaxies, those with M$_{*} >
10^{10}$ \solm at $z \sim 3$, appear to evolve into galaxies with 
stellar masses
$\sim 10^{11} - 10^{12}$ \solm by $z \sim 0$. Thus, from the merger 
process, Lyman-break
galaxies can become the most massive galaxies in today's universe.
The lower stellar mass systems, with M$_{*} > 10^{9}$
and M$_{*} > 10^{8}$ \solm, also increase by up to two orders of magnitude
in stellar mass due to the major merger process (Figure~7).

The majority of this increase in stellar 
mass is due to the major merging activity. 
That is, their stellar mass increases over time because there are on 
average 4-6 major
mergers occurring per galaxy, and each one will slightly less than
double the stellar mass.  The star formation appears to contribute a small fraction, roughly
10-30\% of the new mass, although this can be much higher (up to 50\%) if we 
consider a larger amount of star formation within bursts, as discussed above.

In summary, it appears that
the most massive LBGs at $z \sim 3$ can become galaxies with stellar
masses $\sim
10^{12}$ \solm.  These galaxies rapidly merge between $z \sim 3$ and
$z \sim 2$ where the mass increases by a factor of ten.  We find that 
there are no
mergers, and thus no addition of stellar mass due to the major merger
process at $z < 1.5$, for these systems.
We see a slightly different pattern for the lower mass systems
with M$_{*} > 10^{9}$ \solm and M$_{*} > 10^{8}$ \solm.  Figure~6 shows
that there is not as much merger activity for the lower mass systems
between $z \sim 2$ and $z \sim 3$.  There is however a similar
number, if not more mergers, at lower redshifts. In fact, most of the merging
activity at lower redshift occurs for these lower mass galaxies, consistent
with direct mass determinations of galaxies which are merging at $z < 1$
(Bundy, Ellis \& Conselice 2005).

\section{Discussion}

\subsection{Possible Objections to the Merger Scenario}

We find that massive Lyman-break galaxies at $z \sim 3$ 
will undergo four to six major
mergers from $z \sim 3$ until $z \sim 0.5$, with most occurring
between $z = 2$ and $z = 3$. 
Lower-mass systems have a merger history whereby more mergers occur
at lower redshifts, although these systems also gain up to 100 times
their initial mass through the merger process.

There are several possible objections to this picture, including the number
counts of galaxies at different redshifts, and the observed age of
starbursts in $z > 2$ galaxies. The first requires that the
co-moving number densities of evolved massive LBGs
 match the number density of $z \sim
0$ massive systems.  Since the density of LBGs roughly matches
the density of massive spheroids today (Steidel et al. 1996), we 
cannot destroy, or create
too many new massive galaxies through mergers.  There must be a balance
such that as LBGs effectively disappear through mergers, there must be
fainter and lower mass systems to replace them.  
For the most massive galaxies, this effectively requires in general that
the number of massive galaxies,

$${\rm N} ({\rm M_{*}} > 10^{10} M_{\odot})(z - \delta z) = {\rm N}_{\rm tot}({\rm M_{*}} > 10^{10} M_{\odot})(z)$$
$$- f_{\rm gm}(z) \times {\rm N}_{\rm tot}({\rm M_{*}} > 10^{10} M_{\odot})(z)$$ $$+ f'_{\rm gm}(z) \times {\rm N}_{\rm tot}({\rm M_{*}} > 10^{10-\delta} M_{\odot})(z),$$ 

\noindent remain constant at redshifts $z < 3$, where $\delta$ is
small ($< 1$).  This requires
that $f_{\rm gm}$ $\times$ N$_{\rm tot}$ 
(M$_{\star}$ $> 10^{10}$ \solm) = $f'_{\rm gm}$ $\times$ N$_{\rm tot}$ 
(M$_{\star}$ $> 10^{10-\delta}$ \solm) at all redshifts, or $f_{\rm gm}$/$f'_{\rm gm}$ =
 N$_{\rm tot}$ (M$_{\star}$ $> 10^{10-\delta}$ \solm) /  N$_{\rm tot}$ 
(M$_{\star}$ $> 10^{10}$ \solm).  The value of $f_{\rm gm}$/$f'_{\rm gm}$ $\sim$ 5 at 
$z > 2$, using the 10$^{9.5}$ \solm value for $f'_{\rm gm}$.  This 
is similar to N$_{\rm tot}$ (M$_{\star}$ $> 10^{9.5}$ \solm) /  N$_{\rm tot}$ 
(M$_{\star}$ $> 10^{10}$ \solm) at the same redshift. Based on this, it 
appears 
that the number densities of the most massive galaxies will remain roughly the 
same as the observed densities of LBGs, which are similar to the densities
of modern massive galaxies. This result originates from the fact
that while there is a higher
merger fraction for the more massive galaxies, there are more lower mass
galaxies undergoing mergers at all redshifts (Figure~5).

A related issue is the stellar mass to dark matter ratio of Lyman-break
galaxies.  Although this ratio is not know for certain, there are indications
that the Lyman-break galaxies are associated with the most massive 
halos of today (Adelberger et al. 1998) while stellar masses for the most
massive systems are typically around 10$^{10}$ \solm.  If mergers are the
dominate source of the build up of galaxies from LBGs without any star
formation, the ratio of stellar to total mass would remain constant. If the
halo masses of LBGs are of order 10$^{12}$ \solm, then there must be
further star formation to match the ratio of stellar to halo masses
of 0.1 found today and even by $z \sim 1$ (Conselice et al. 2005).  
Although we find that up to 30\% of the addition mass is formed in
star formation, the majority originates from the existing mass in the
mergers. It is also not certain that LBGs have a stellar to total
mass ratio of 0.01.  Often, the total masses of $z > 2$ galaxies
are less than their stellar masses (Shapely et al. 2004), making
the total or stellar (but likely total) masses suspect.  Furthermore, the
halo occupation number for LBG halos may be larger than one, such that there
are more than one galaxy in a single massive halos.  This indeed seems
to be the case based on recent measurements of the LBG correlation function
which show that there is an excess at small scales (Lee et al. 2005) - 
possibly the result of ongoing mergers and multiple galaxies in single halos.

Another possible objection is that the ages of the stellar populations 
in LBGs are too old at $z > 2$ to be produced in bursts induced by 
the large number of mergers occurring
between $z \sim 2$ and
$z \sim 3$.  In other words, the observed ages of some starbursts
between $z \sim 2$ and $z \sim 3$ is longer than the elapsed time
between $z \sim 2$ and $z \sim3$, and thus it would be difficult to have
several mergers, and accompanying star formation occur during this time.
Papovich et al. (2001) and Shapley et al.
(2001) have found a wide range of stellar population or burst ages
for galaxies within this redshifts range.  Some of these ages are as
old as 1 Gyr.  Clearly
not all LBGs are involved in the multiple mergers discussed here unless
these time estimates are incorrect.  There is some evidence for this
as the age of the starburst can very by many factors depending
on the choice of IMF and metallicity (Papovich et al. 2001).  However,
there are reasons to believe these 'older' bursts are occurring in the lower
mass systems (Shapley et al. 2001) which are also the more symmetric systems
(Conselice et al. 2003a,c).

In general however, it appears that many of the best fit starburst ages for 
LBGs are short $< 100$ Myrs, particularly when a metallicity of 
0.2 solar is used (Papovich et al. 2001).  In fact, the systems identified
in Conselice et al. (2003a), and in this paper, as mergers have young starburst
ages,  consistent with interpreting their star formation
produced in a recent merger event.  This correlation can be seen in 
Figure~8 which plots the ages of the most recent starburst, and the
asymmetries of their galaxies, for $z > 2$ LBGs (Papovich et al. 2001).  
The galaxies with high asymmetries, $A > 0.35$, and thus likely merging 
(above the solid line), all
have young ages, typically less than 50 Myrs.  These merging systems
also have the highest ongoing star formation rates.  Since N-body models
show that not all phases of a merger have a high asymmetry, it is possible
that the symmetric young age LBGs are in a phase of a merger where they are
not asymmetric.

\subsection{Implications for Massive Galaxy Formation}

If the merger history at $z > 1$ as derived in Conselice et al. (2003a), and
used in this paper to calculate the formation history of galaxies holds
up with future observations and techniques it has profound implications.
It implies that we have observationally solved how and when
most massive galaxies formed.  Our results are qualitatively consistent with 
several
other apparently paradoxically results, a number of which have
questioned the foundation of the modern theoretical galaxy formation
paradigm, Cold Dark Matter.

Recently various groups have discovered
massive galaxies at redshifts $z > 1$, which has been seen
as potentially a problem for CDM based models (Franx et al. 2003; Daddi
et al. 2004; Somerville et al. 2004;
Glazebrook et al. 2004; cf. Nagamine et al. 2005). 
For example Glazebrook et al. (2004) find that the stellar mass
density of massive galaxies, with M$_{*} > 10^{10.8}$ \solm, is roughly an
order of magnitude larger than expectations based on a semi-analytic CDM 
model of galaxy formation (Cole et al. 2000; Benson et al. 2002), 
although the
agreement with the M$_{*} > 10^{10.2}$ \solm models is quite good.  This
implies that the most massive galaxies formed earlier than what CDM
models predict.  There are two possible solutions to this. The first, presented
in this paper, is that the merger rate, and the fraction of galaxies at
high redshift which are merging, is higher than what is predicted in  
CDM models.      

This mismatch with CDM models can be seen through the
comparison in
Figure~1. Figure~1 shows that this particular CDM model, from
semi-analytic galaxy formation modeling  
(Benson et al. 2002), generally underpredict the merger
fraction we observe, as do other semi-analytical model
predictions (Somerville et al. 2001).  This suggests that the solution to 
the `massive 
galaxies at high redshift problem' is not a rapid collapse with
no mergers, but that there are possibly {\em more} mergers at higher redshift
than what semi-analytic CDM models predict.  Our results suggest that
massive galaxies should be well formed by $z \sim 1.5$, which is consistent
with observations of galaxy stellar masses (Figure~7).  
An alternative scenario is that these studies,
 are biased by cosmic variance. It
is possible that the Glazebrook et al. (2004) and Franx et al. (2003) studies
are examining over dense regions at their particular redshifts.  Wider
area infrared surveys  are clearly
needed to make progress in understanding this issue.

An early formation history for massive early-type galaxies explains a number
of other observations.  The first is that the morphological distribution of
galaxies on the Hubble sequence is largely in place by $z \sim 1$ (Conselice
et al. 2005), with a similar number density of disk galaxies as today
(Ravindranath et al.
2004).  Another is that the clustering
properties of massive galaxies appear to be largely in place by $z \sim 1$
(Coil et al. 2004), as is the size distribution (Ferguson et al. 2004),
with galaxy sizes growing between $2 < z < 6$ (Bouwens et al.
2004).   The metallicities of massive galaxies at $z > 2$ are also
similar to the most massive galaxies found today, with values
of solar or greater (Shapley et al. 2004; van Dokkum et a. 2004), suggesting
that there are not many future generations of massive starbursts.   
Another observation consistent with a rapid and early
merging history for massive galaxies is that the number of metal poor
globular clusters around early-types correlates with 
galaxy luminosity (Strader et al. 2004), implying that their formation  
occurred at $z > 2$.    This does not however relieve
other problems with the hierarchical formation of galaxies, as there
are massive systems well formed by $z \sim 6$ in the form of
QSOs that contain high metallicities (Barth et al. 2003) 
that might also contain massive galaxies.  Clearly
probing the merging history at $z > 3$ will be insightful.  

Major mergers at $z < 3$ also  relate to the build
up of black holes at similar redshifts.  The peak of the merger rate and
merger fraction coincides with the peak of AGN and QSO activity
(Boyle et al. 2000).  Thus it seems likely that the build up of black
holes and galaxy bulges, which is already present in some form at
$z \sim 1$ (Grogin et al. 2005) could be driven by the major merger process.
As there is little correlation between the presence of merging activity
and X-ray flux for LBGs (Lehmer et al. 2005), this implies that there is
some delay between the formation of black holes and the merger.

One problem that we have not addressed is the fact that only $\sim$ 10-20\% of
all stellar mass is formed by $z \sim 3$ (Dickinson et al. 2003; Rudnick et al.
2003), and thus there must be, and is, significant star formation at
$z < 3$ (Giavalisco et al. 2004b).  Since only 10-30\% of the stars
in a massive galaxy form by starbursts induced in major mergers, 
there must be other methods of forming new
stars at $z < 3$.   At lower redshifts when 
the merger rate declines, other methods such as dissipation to form disks at
$1 < z < 2$, or secular evolution produced by bars which are common
at $z \sim 1$ (Jogee et al. 2004), must dominate the stellar mass assembly.  
Since spheroids dominate the stellar mass density
of nearby galaxies, containing roughly a third to half of all stars (Tasca 
\& White 2005), it is not likely that all this mass is
produced in major mergers. It appears that the lower
mass spheroids are produced through dissipative processes at $z < 1$
when they are found to be bluer systems (Stanford et al. 2004), while
the higher mass systems, which are generally in denser environments,
are produced through major mergers at earlier times.

We have also not discussed the relationship between the sub-mm bright
galaxies found at $z > 1$ and the systems studied in this paper.  The
Lyman-break galaxies we examined are much more common at high
redshift than sub-mm sources, yet the relationship between 
sub-mm sources and Lyman-break galaxies is still largely unknown.
It is possible that the sub-mm sources are the most massive galaxies
at high redshift, as suggested by their clustering properties (Blain
et al. 2004).   In any case,
the morphologies of the limited number of sub-mm sources that have
been studied show that these systems are large galaxies that
are undergoing major mergers (Conselice et al. 2003c; Chapman et al.
2003; Pope et al. 2005), with preliminary indications that the sub-mm sources
are involved in a more active phase of a merger than the most massive
LBGs studied in the Hubble Deep Field (Conselice et al. 2003b).

\section{Summary}

In the first part of this paper we reduce all known merger fractions
to a common scale based on the fraction of {\em galaxies} undergoing
a merger at a given redshift, and within a range of stellar
mass or luminosity.  We then show that the best fitting function for
the galaxy merger fraction history is a combined power/exponential formalism 
for systems
with M$_{*} < 10^{10}$ \solm and M$_{\rm B} > -20$ out to $z \sim 3$. 
For the brightest and most massive systems with M$_{\rm B} < -20$ and
M$_{*} > 10^{10}$ \solm out to $z \sim 3$, and for lower mass and fainter
systems out to $z \sim 1$, a simple power-law is suitable for describing
the redshift dependence of the merger fraction.

We then analyze a suite of self-consistent major and minor merger N-body 
simulations
with the CAS structural analysis system.  We find that major mergers,
defined as having a mass ratio of 1:3 or lower, always produce
asymmetric systems, typically for 300 Myrs, independent of viewing 
angle or relative orbital configuration of the pair, with a slight
dependence on mass which we account for.  We investigate
in detail the structure of a 1:5 merger and find that during the 4 Gyr
duration of this merger, the asymmetry parameter never becomes high enough to
be identified as a merger in the CAS system.  
We then use this information to derive
the galaxy merger rate.    
From this we calculate that an average massive galaxy,
with M$_{*} > 10^{10}$ \solm, undergoes 4.4$^{+1.6}_{-0.9}$ major mergers 
using our derived merger time-scale that slightly depends on mass (eq. 10).  
Nearly all of this
merging occurs by $z \sim 1.5$, after which an average massive
galaxy experiences no further major mergers.

We then calculate  how
galaxies of various initial masses at $z \sim 3$ formed through major 
mergers.  Our conclusion is that mass accreted during the major merger process,
and the stars created from star formation induced by these mergers, is such 
that when these systems evolve to $z \sim 0$ they will be as massive as the 
most galaxies in the modern universe.  We calculate that a typical galaxy 
increases in mass by 
a factor of $\sim 10-100$ from this process.  
Our calculations are based on empirically
determined merger fractions and rates from Conselice et al. (2003a) and
include star formation scenarios based on observations of
star formation histories of Lyman-Break galaxies.  We calculate that
10-30\% of the new stellar mass is formed in starbursts induced by these
mergers, and the remainder comes from the merger itself.   The merger scenario
described in this paper naturally
explains observations of massive and extremely red galaxies 
at $z > 1$, the distributions of modern 
elliptical galaxy ages and metallicities, among many other  
massive galaxy properties.

Future results will expand our conclusions using better models
and deeper HST observations.    Wider
area and deep infrared surveys at high resolution, either with WFC3 on HST,
with JWST, and/or with ground based adaptive optics, are needed to determine
with certainty the merger rate at high redshift. ACS imaging of
GOODS fields (Giavalisco et al. 2004a) will potentially
allow us to make these determinations, although high-resolution
deep infrared imaging is needed to make definitive progress. 

\vspace{1cm}

The idea for this paper was initiated by a question from
Harry Ferguson.  Although, as Jim Gunn says, ``Partial answers are
the only answers'' (Morgan 1988), and unfortunately this work is no exception.
I hope, at the very least, that this presentation begins a serious discussion 
of determining observationally {\em how} galaxies formed.  I thank Chris Mihos 
for the use of his models, and his critically important contributions to 
this paper, and for being a true gentleman. This work has furthermore 
benefited from collaborations and conversations
with Matt Bershady, Kevin Bundy, Mark Dickinson, Richard Ellis, Jay
Gallagher and Casey Papovich.  I furthermore thank Xavier Hernandez and James
Taylor for illuminating discussions, Kevin Bundy and Russel White for 
helpful proofreads, Andrew Benson for the 
Galform models used in this paper, and finally the referee for making
several important points.  Support for this research was provided by
NSF Astronomy \& Astrophysics Postdoctoral Fellowship \#0201656.

\clearpage

\begin{inlinefigure}
\begin{center}
\vspace{2cm}
\hspace{-0.5cm}
\rotatebox{0}{
\resizebox{\textwidth}{!}{\includegraphics[bb = 25 25 625 625]{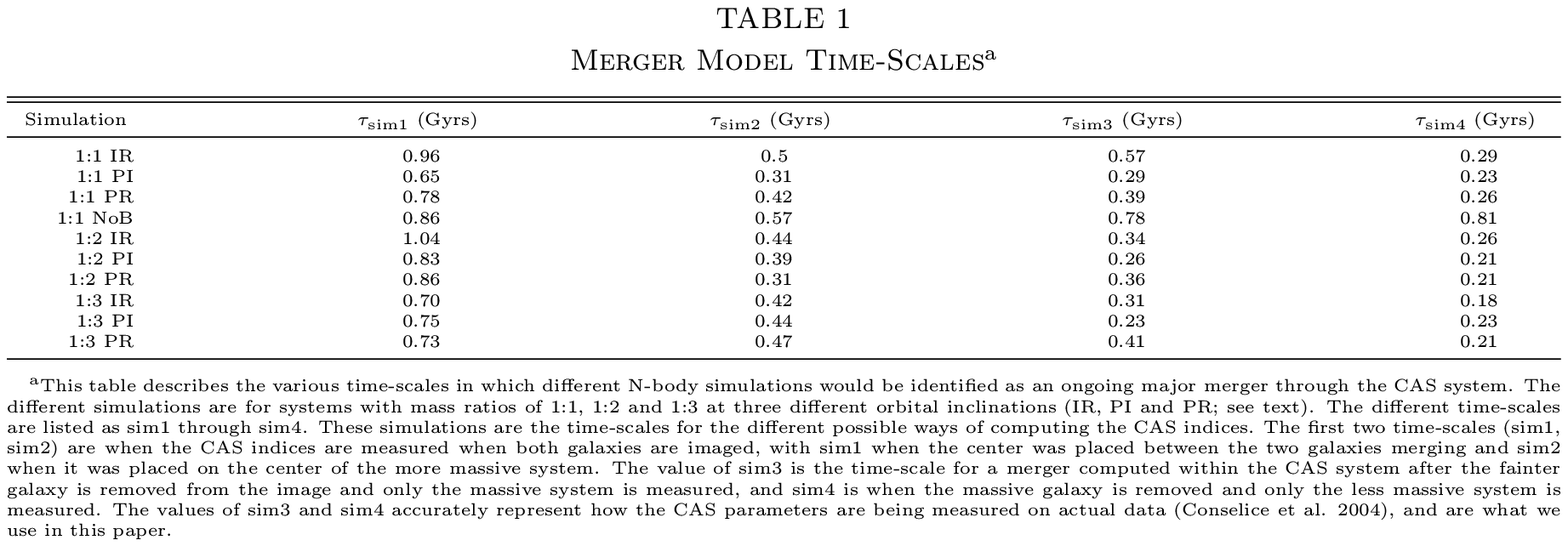}}}
\end{center}
\vspace{-2cm}
\end{inlinefigure}

\clearpage

\begin{figure*}
\begin{center}
\vspace{0cm}
\hspace{-1cm}
\rotatebox{0}{
\includegraphics[width=1\linewidth]{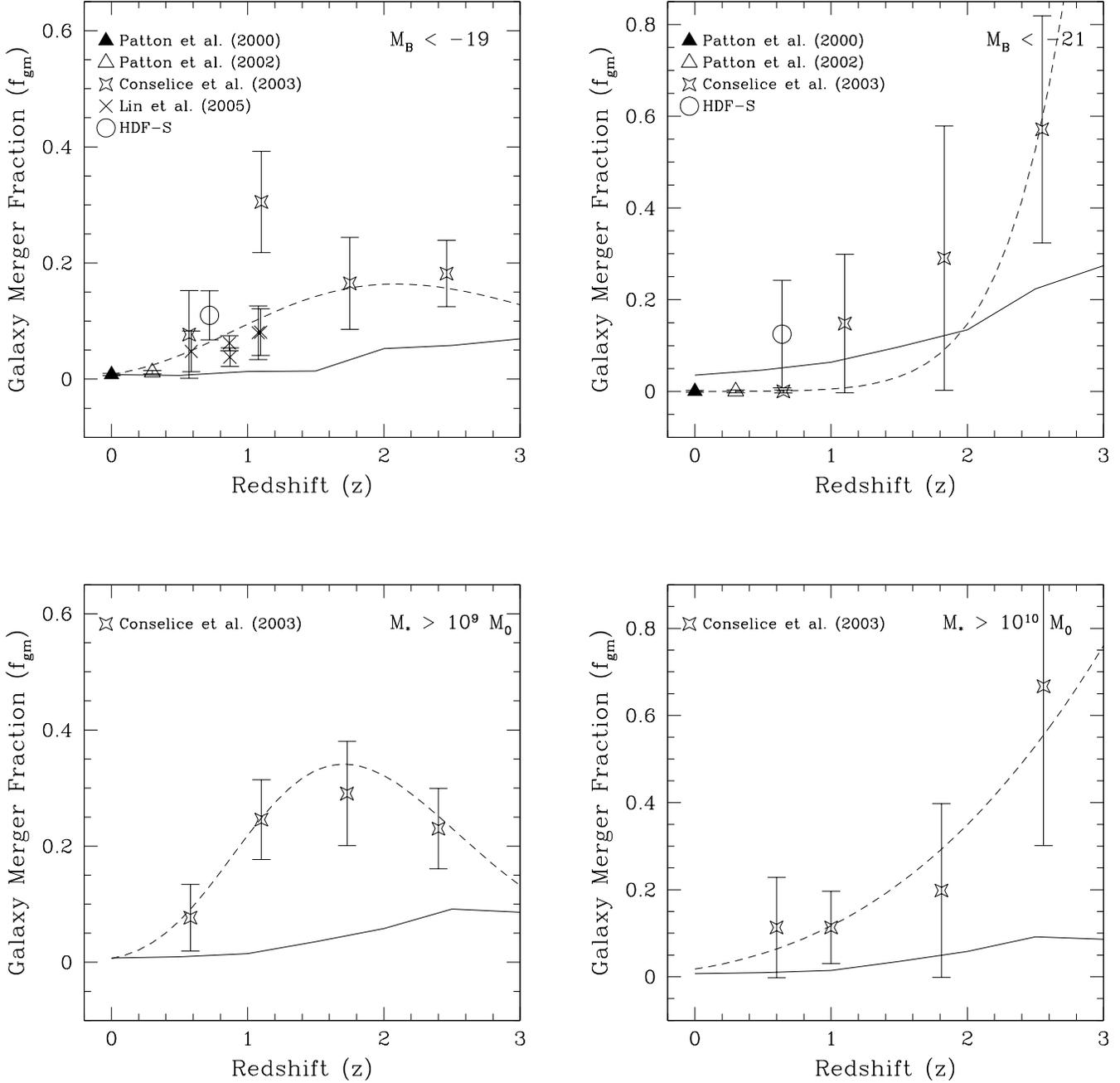}}
\end{center}
\vspace{-0.5cm}
\figcaption{The merger fraction at two different luminosity limits, M$_{\rm B} 
< -19$ (upper left hand panel) and M$_{\rm B} < -21$ (upper right panel) and
two upper mass limits, M$_{*} > 10^{9}$ \solm (lower left panel) and M$_{*} > 
10^{10}$ \solm (lower right panel).  Plotted
are merger fractions from pair counts and morphological
measures from Patton et al. (2000),
Patton et al. (2002), Conselice et al. (2003) and Lin et al. (2005). The
dashed lines are the best fit to the merger fractions, where a power-law
of the form $f_{\rm gm} = f_{0} \times (1+z)^{\rm m}$ is fit for the
systems with M$_{\rm B} < -21$ and M$_{*} > 10^{10}$ \solm and a 
power-law/exponential of the form $f_{\rm gm} = \alpha (1+z)^{\rm m} \times {\rm exp}(\beta(1+z))$ is fit for systems with  M$_{\rm B} < -19$ and 
M$_{*} > 10^{9}$ \solm.  The  solid
line shows CDM model predictions for mergers occurring within
these luminosities and stellar masses.}
\end{figure*}

\clearpage

\begin{figure*}
\begin{center}
\vspace{0cm}
\hspace{-1cm}
\rotatebox{0}{
\includegraphics[width=1\linewidth]{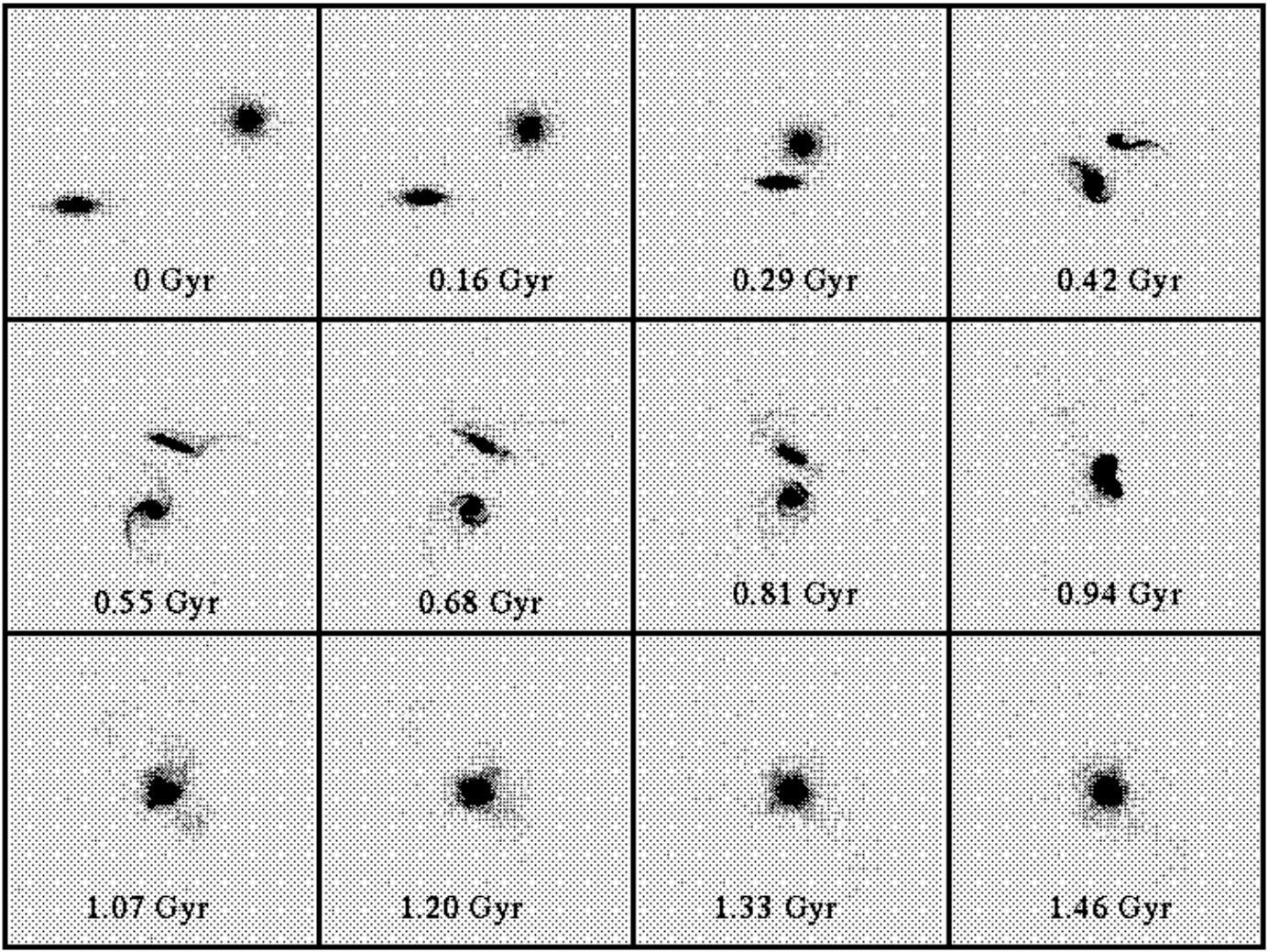}}
\end{center}
\vspace{-0.5cm}
\figcaption{A visual realization of a 1:1 merger simulation analyzed in
this paper. The time at the bottom of each frame is the time during
the simulation up until 1.46 Gyr.}
\end{figure*}

\clearpage

\begin{figure*}
\begin{center}
\vspace{0cm}
\hspace{-1cm}
\rotatebox{0}{
\includegraphics[width=1\linewidth]{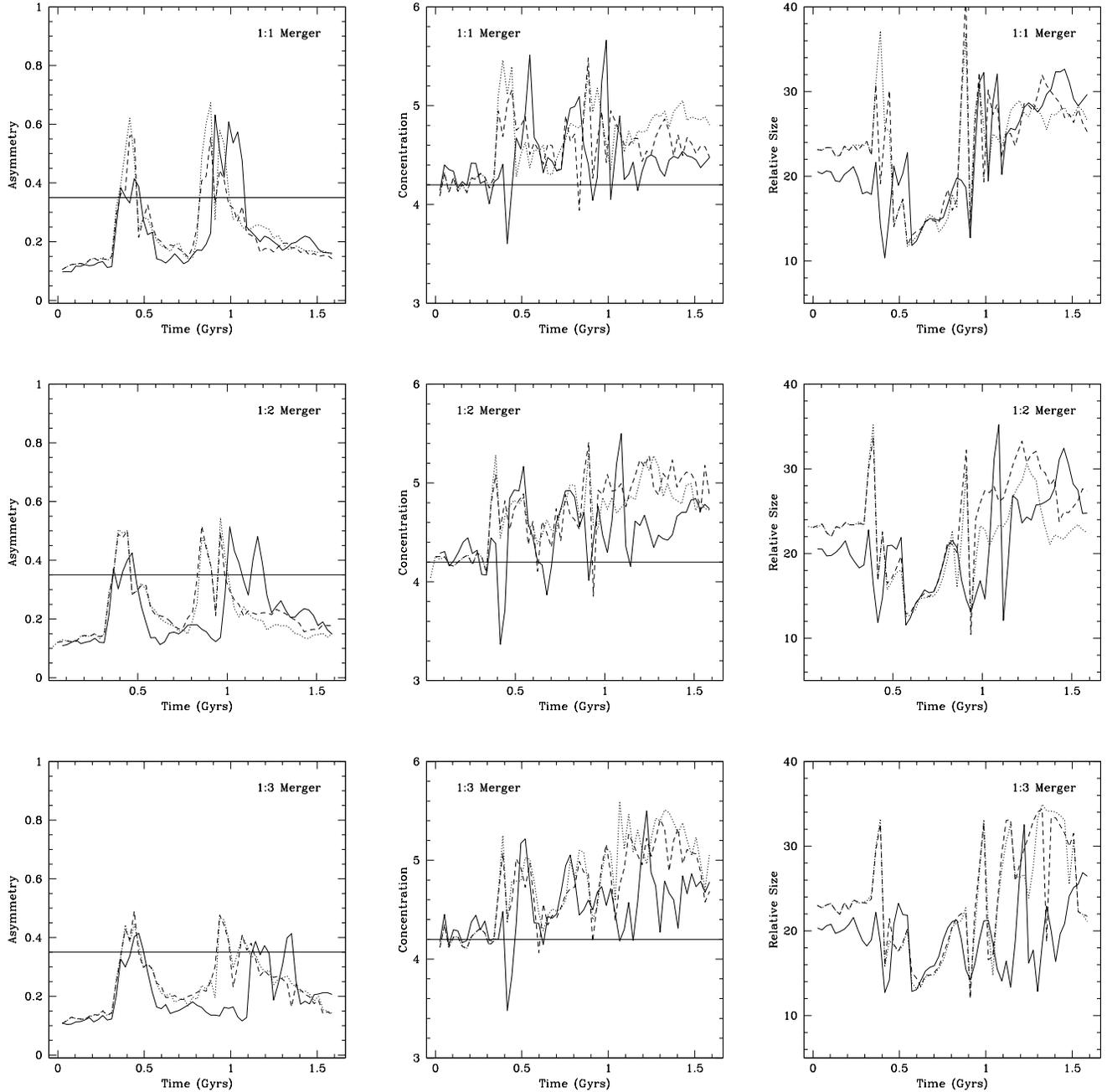}}
\end{center}
\vspace{-0.5cm}
\figcaption{The evolution of the asymmetry, concentration and sizes for 
galaxies in our N-body simulations.  The three dashed lines are for simulations
with different types of orbits, with the IR (solid), PI (dashed) and 
PR (dotted) models shown.  The solid line in the asymmetry
diagram is the limit for finding galaxy mergers (Conselice 2003).  The solid
line in the concentration diagram shows an average value for bulge
dominated galaxies.}
\end{figure*}

\clearpage

\begin{figure*}
\begin{center}
\vspace{0cm}
\hspace{-1cm}
\rotatebox{0}{
\includegraphics[width=1\linewidth]{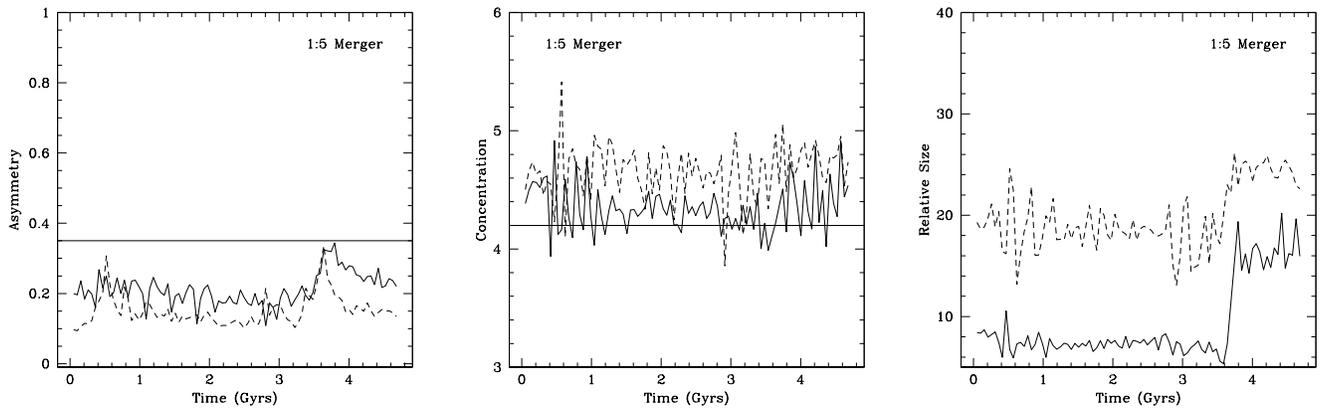}}
\end{center}
\vspace{-0.5cm}
\figcaption{Similar to Figure~3, but for mergers that have a mass
ratio of 1:5.  The lines are the same as in Figure~3, except that
in this case the two lines that represent the simulations 
(solid and dashed) are the asymmetry (A), concentration (C), and size
values for the simulated (less and more massive of the pair) galaxies.}
\end{figure*}

\clearpage

\begin{figure*}
\begin{center}
\vspace{0cm}
\hspace{-1cm}
\rotatebox{0}{
\includegraphics[width=1\linewidth]{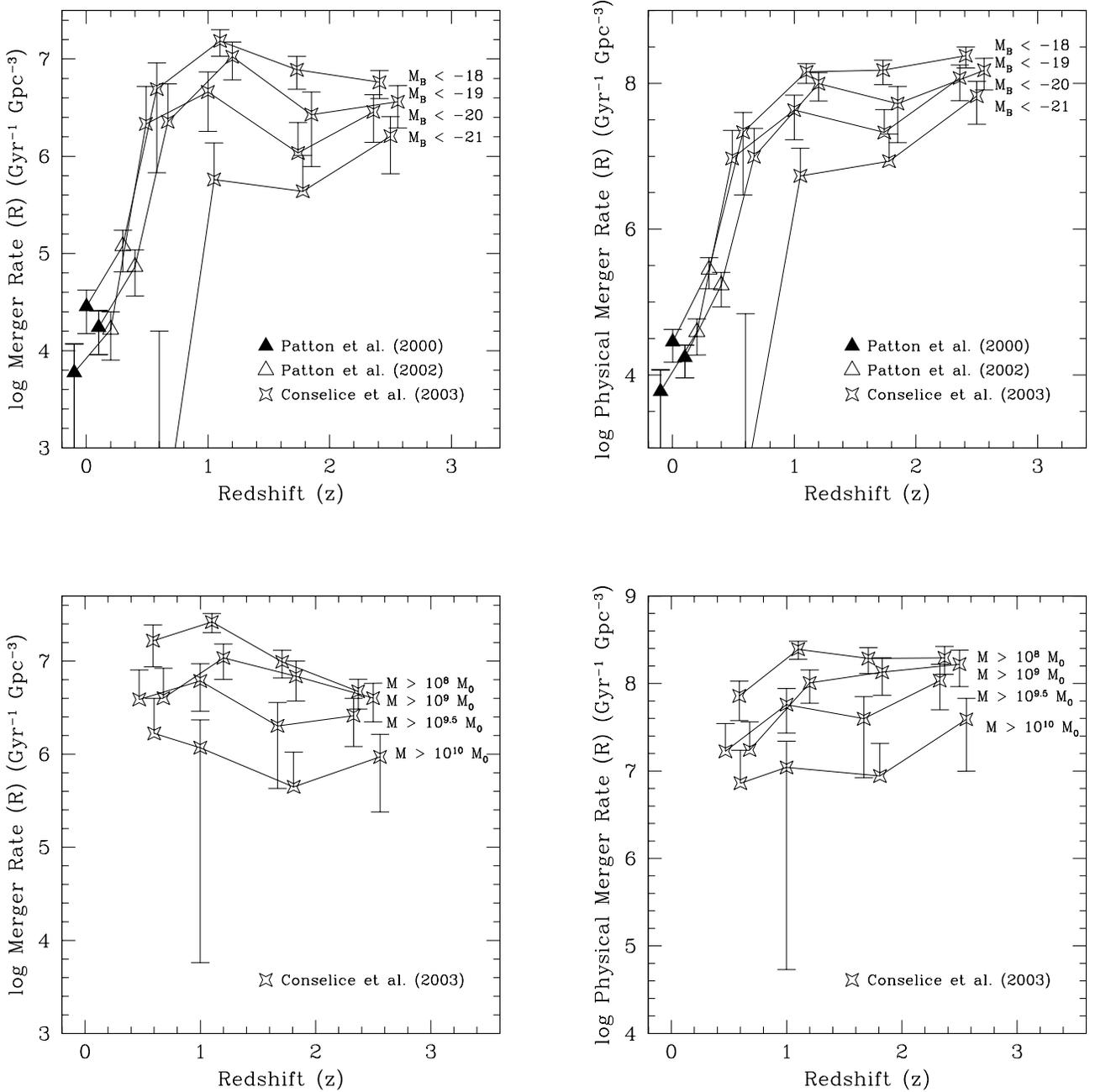}}
\end{center}
\vspace{-0.5cm}
\figcaption{The history of galaxy merger rates, in units of Gyr and Gpc$^{3}$,
as a function of different 
initial masses and luminosities starting at $z \sim 3$.  
Plotted on the left is the merger rate
within co-moving volumes, while the right panels plot the merger rate 
within physical volumes.  Plotted here are merger rates computed
from the observed galaxy merger fractions in Conselice et al. (2003), Patton
et al. (2002) and Patton et al. (2000).  Some points are shifted
by $\pm0.1$ in redshift to allow individual points to be better seen.  
The error
bars include uncertainties from the merger pair fractions as well as 
uncertainties from the merger time-scales.  Errors which are larger
than their data point are shown by (accurate) one sided upper error 
limits.  The M$_{\rm B} < -21$ point,
below the 3 Gyr$^{-1}$ Gpc$^{-3}$ plot limit on the co-moving merger rate
plot has a value of 1.9 Gyr$^{-1}$ Gpc$^{-3}$ at $z = 0.6$. Likewise, the
similar point in the physical merger rate has a value 2.5 Gyr$^{-1}$ Gpc$^{-3}$at $z = 0.6$. }
\end{figure*}

\clearpage

\begin{figure*}
\begin{center}
\vspace{0cm}
\hspace{-1cm}
\rotatebox{0}{
\includegraphics[width=1\linewidth]{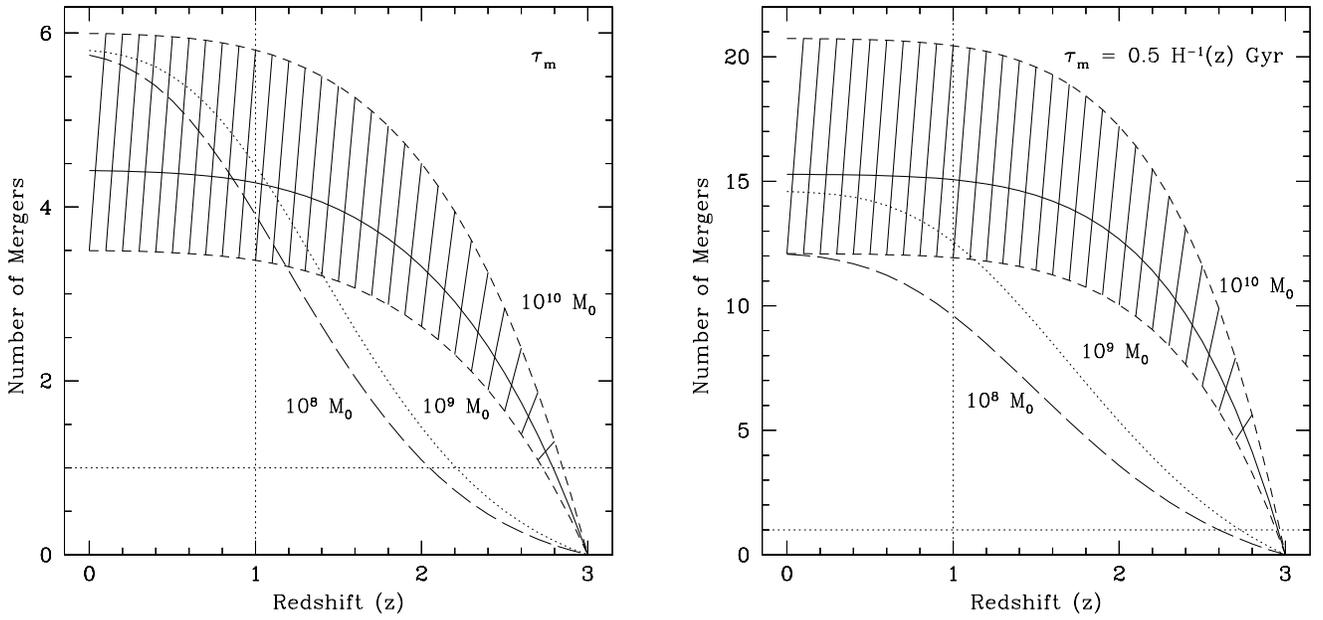}}
\end{center}
\vspace{-0.5cm}
\figcaption{The cumulative number of major mergers (1:3 or lower mass ratios) 
for galaxies starting 
from $z = 3$ for systems with different initial stellar masses. The solid
line shows the evolution in the cumulative number of mergers for an 
average galaxy with an initial mass of 10$^{10}$ \solm, with the hatched region
showing the 1 $\sigma$ range of possible outcomes. This 1 $\sigma$
range includes uncertainties in the merger time scales as well as 
uncertainties in the measured merger fractions.
 For lower-mass galaxies, the number of mergers is shown
as the dashed line for systems with M$_{\star} > 10^{8}$ \solm  and
the dotted line for systems with M$_{\star} > 10^{9}$ \solm.}
\end{figure*}

\clearpage

\begin{figure*}
\begin{center}
\vspace{0cm}
\hspace{-1cm}
\rotatebox{0}{
\includegraphics[width=1\linewidth]{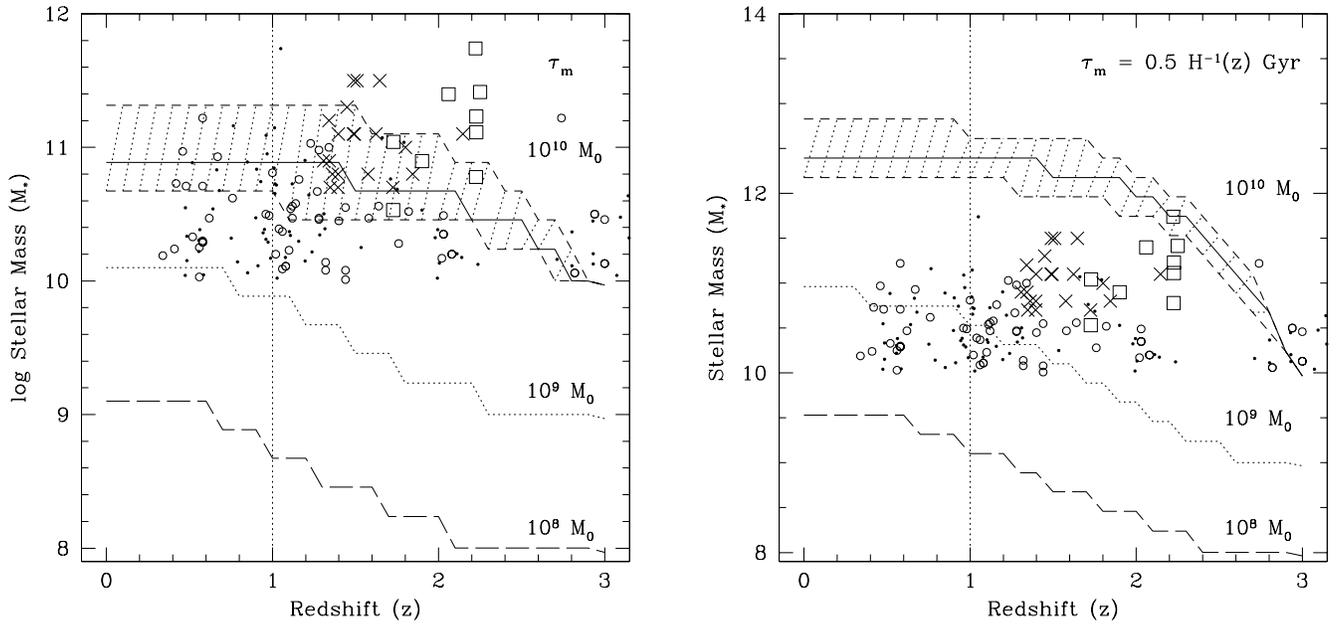}}
\end{center}
\vspace{-0.5cm}
\figcaption{The average evolution of stellar masses for 
galaxies with various initial stellar
masses, starting at: M$_{\star}$ = 10$^{10}$ \solm, 
M$_{\star}$ = 10$^{9}$ \solm, and M$_{\star}$ = 10$^{8}$ \solm.  The 
vertical dotted line is
$z \sim 1$.  The range of possible final stellar masses using the
observed merger fraction evolution, and the amount of induced
stellar mass, is shown in the shaded region.  Also
plotted are observed stellar masses for galaxies from the HDF-N (Dickinson
et al. 2003; dots), HDF-South (Franx et al. 2003; Conselice et al. 2005;
open circles), K20 survey (Daddi et al. 2004; boxes) and the Gemini
Deep Deep Survey (McCarthy et al. 2004; crosses).}
\end{figure*}

\clearpage

\begin{inlinefigure}
\begin{center}
\vspace{2cm}
\hspace{-0.5cm}
\rotatebox{0}{
\resizebox{\textwidth}{!}{\includegraphics[bb = 25 25 625 625]{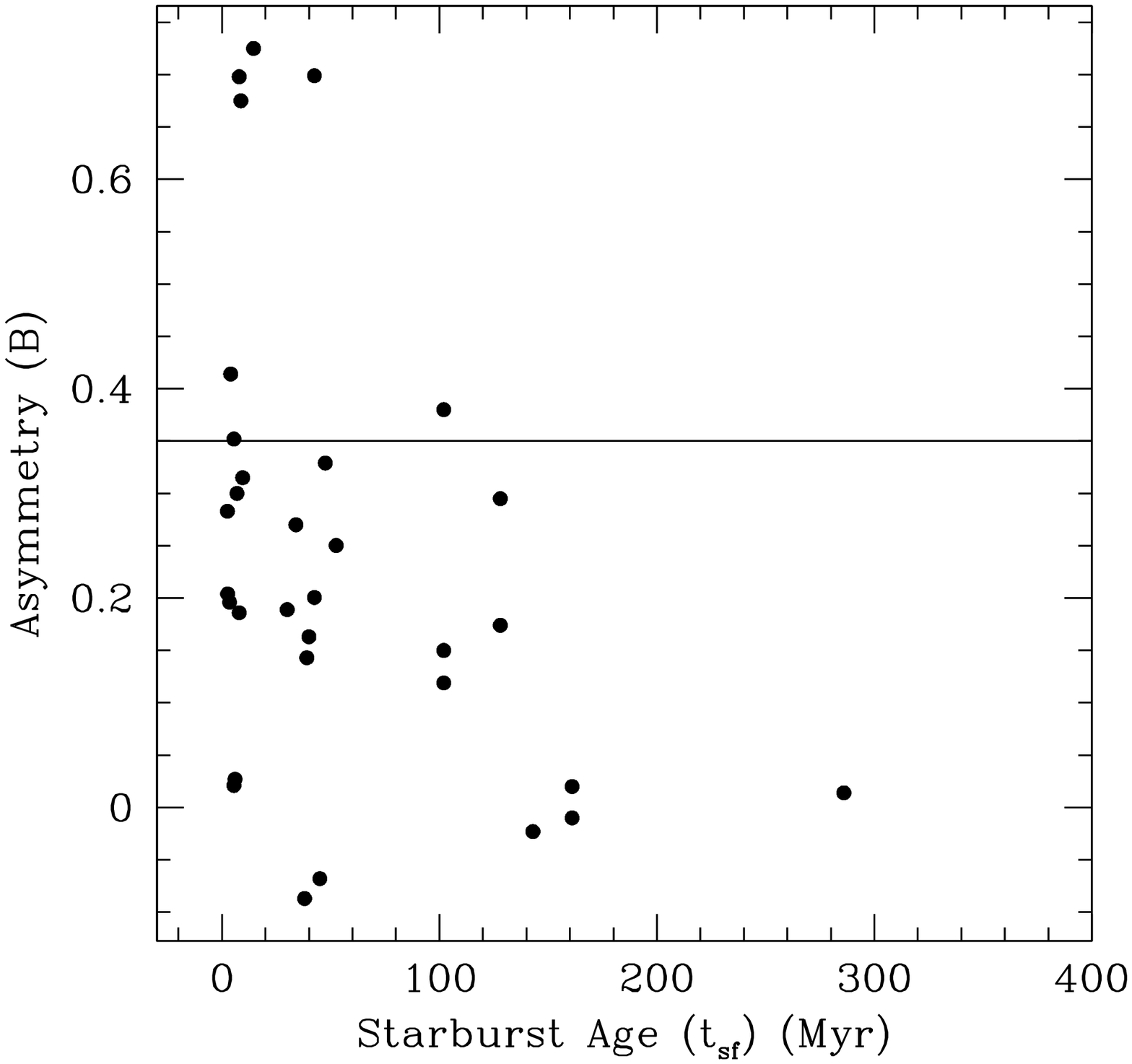}}}
\end{center}
\vspace{-2cm}
\figcaption{The relationship between the age of the most recent starburst for
galaxies at $z > 2$, as a function of rest-frame B-band
asymmetry values. These
ages are taken from Papovich et al. (2001) utilizing the 0.2 solar and
Salpeter IMF models.  The
systems with asymmetries consistent with undergoing a merger are above
the solid line. These systems typically have younger starburst ages. 
This implies that the asymmetry index is a good indicator for recent
mergers, as all of these systems have ages $< 100$ Myrs. }
\end{inlinefigure}

\end{document}